\documentclass[showkeys,preprint,aps,prd,showpacs,preprintnumbers,amsmath,amssymb,amsfonts, nofootinbib,a4paper,12pt]{revtex4-1}
\usepackage{hyperref}
 \usepackage{enumerate}
 \usepackage{graphicx} 
 \usepackage{amsmath}
\usepackage{amssymb}
\usepackage{graphicx}
\usepackage{epstopdf}
 \numberwithin{equation}{section} 
\usepackage{bbm,amsmath,graphicx,amssymb,amsfonts,amsthm} 
\usepackage{bm} 
\usepackage[T1]{fontenc}
\usepackage[latin1]{inputenc}
\usepackage{t1enc}
\usepackage{graphicx}
\usepackage{amssymb}
\usepackage{amsmath}

\usepackage{float} 
\usepackage{mathtools}  
\usepackage{blkarray}

\usepackage{verbatim} 
\usepackage{latexsym} 
\usepackage{tikz}
\usepackage[english]{babel}

\def\be{\begin{equation}}
\def\ee{\end{equation}}
\def\ba{\begin{eqnarray}}
\def\ea{\end{eqnarray}}

\def\a{\alpha}

\def\l{\langle}

\def\CP1{\mathbb{CP}^1}
\def\SL2C{\mathrm{SL}(2,\mathbb{C})}

\def\Z2{\mathbb{Z}_2}

\def\su2{{SU(2)}}
\def\eps{{\epsilon}}
\def\a{{\alpha}}

\def\sz{{\it z}}
\def\[{\left[}
\def\]{\right]}

\def\l{\lambda}

\def\a{\alpha}

\def\({\left(}
\def\){\right)}
\def\[{\left[}
\def\]{\right]}

\def\<{\langle}
\def\>{\rangle}

\def\i2{\frac{i}{2}}

\def\2F1{\,_2{\rm F}_1}

\newcommand{\beq}{\begin{equation}}
\newcommand{\eeq}{\end{equation}}
\newcommand{\beqq}{\begin{equation*}}
\newcommand{\eeqq}{\end{equation*}}
\newcommand\beqa{\begin{eqnarray}}
\newcommand\eeqa{\end{eqnarray}}
\newcommand\beqaa{\begin{eqnarray*}}
\newcommand\eeqaa{\end{eqnarray*}}
\newcommand\bea{\begin{array}}
\newcommand\eea{\end{array}}
\newcommand{\ie}{{\it i.e.}}

\newcommand{\sket}[1]{\left|#1\right\rbrack}
\newcommand{\sbra}[1]{\left\lbrack#1\right|}
\newcommand{\sbraket}[1]{\left\lbrack#1\right\rbrack}
\newcommand{\ket}[1]{\left|#1\right\rangle}

\newcommand{\braket}[1]{\left\langle#1\right\rangle}
\newcommand\Tstrut{\rule{0pt}{2.6ex}}     
\newcommand\Bstrut{\rule[-0.9ex]{0pt}{0pt}}  

\renewcommand{\baselinestretch}{1.5}
\begin{document}
\title{\Large One-Loop Yang-Mills Integrands from Scattering Equations}
\renewcommand{\baselinestretch}{1}
\author{\vspace{0.9cm}{\bf
Johannes Agerskov${}^{a,b}$, 
N.~E.~J.~Bjerrum-Bohr${}^{b}$, 
Humberto Gomez$^{b,c}$ and  
Cristhiam Lopez-Arcos$^{d}$\\ \vskip0.7cm}
${}^a$\,Department of Mathematical Sciences, University of Copenhagen\\
Universitetsparken 5, DK-2100 Copenhagen, Denmark\\[5pt]
${}^{b}$\,Niels Bohr International Academy and Discovery Center\\
The Niels Bohr Institute, University of Copenhagen\\
Blegdamsvej 17, DK-2100 Copenhagen, Denmark\\[5pt]
${}^c$Facultad de Ciencias, Basicas Universidad\\ Santiago de Cali,
Calle 5 $N^\circ$ 62-00 Barrio Pampalinda, Cali, Valle, Colombia\\[5pt]
${}^d$\,Grupo de Electr\'{o}nica y Automatizaci\'{o}n,
Instituci\'{o}n universitaria Salazar y Herrera 
Carrera 70 $N^\circ$ 52-49, Mede$ll\acute{\imath}n$, Colombia\\[15pt]
{\small {\bf Email:} johannes-as@math.ku.dk, bjbohr@nbi.dk, humberto.gomez@nbi.ku.dk, crismalo@gmail.com\bigskip\bigskip\bigskip}
}
\date{\today}

\begin{abstract}\bigskip\bigskip\bigskip
We investigate in the context of the scattering equations, how one-loop 
linear propagator integrands in gauge theories can be linked to integrands with quadratic propagators using a double forward limit. We illustrate our procedure 
through examples and demonstrate how the different parts of the derived quadratic 
integrand are consistent with cut-integrands derived from four-dimensional 
generalized unitarity. We also comment on applications and discuss possible further generalizations.
\end{abstract}
\maketitle
\renewcommand{\baselinestretch}{1.5}
\section{Introduction}\label{Sec.Intro}
The search for computational techniques for scattering amplitudes in quantum field theory is an area under constant development. In the remarkable series of papers by Cachazo, 
He and Yuan (CHY)~\cite{Cachazo:2013gna,Cachazo:2013hca,Cachazo:2013iea} 
it was shown that one can obtain tree-level S-matrix amplitudes in arbitrary dimension for a broad variety of theories, in the context of the so-called \emph{scattering equations}: 
\begin{equation}
S_i\equiv \left(\sum_{j\neq i}\frac{s_{ij}}{z_i-z_j}=0\right),\quad i\in\{1,2,..,n\}\,.
\end{equation}
Here $ s_{ij} \equiv ( k_i+ k_j)^2$ are the usual Mandelstam invariants, defined from the external momenta $k_i$ and $k_j$, and the variables $\sz_i$ and $\sz_j$ are auxiliary coordinates. The scattering equations exhibit $ \text{PSL}(2,\mathbb{C}) $ invariance, and thus only a subset of them are independent. Amplitudes are given by a contour integral enclosing the solutions to the scattering equations,
\ba
A_n \equiv \int_{\Gamma} \,d\mu_n\,\, \cal{I}^{\rm CHY}({\it z})\,, 
\ea
where $ \mathcal{I}^{\text{CHY}}(\sz)$ is the integrand and we employ the integration measure, $d\mu_n$,
\begin{equation}
d\mu_n \equiv\frac{\prod_{a=1}^n d\sz_a}{{\rm Vol}\,({\rm PSL}(2,\mathbb{C}))} \times \frac{(\sz_i-\sz_j) \,(\sz_{j}-\sz_k)\, (\sz_{k}-\sz_i)}{  \prod_{b\neq i,j,k}^{n} S_b }\,.
\end{equation}
Given the tree-level scattering equation formalism, it is natural to speculate about possible loop-level applications. A breakthrough was provided when Geyer, Mason, Monteiro, and Tourkine~\cite{Geyer:2015bja} provided an explicit formalism in the context of ambitwistor string theory~\cite{Mason:2013sva,Adamo:2013tsa}. This was followed by work on one-loop scalar $\phi^3$ theories~\cite{Baadsgaard:2015hia} (see also Ref.~\cite{He:2015yua}) extending the tree-level scattering equation integration rules~\cite{Bjerrum-Bohr:2014qwa}
to loop level. The main idea behind the scattering equation extensions at the one-loop level is the observation that the forward limit of loop amplitudes can be addressed by adding to tree-level scattering processes two additional off-shell loop momenta which are equal and opposite. A two-loop version of this formalism was developed in Refs.~\cite{Feng:2016nrf,Geyer:2018xwu}.\\[5pt]
A feature of the scattering equation formalism is that propagators are obtained linearly instead of quadratically. Quadratic propagators can be decomposed in terms of linear propagators through partial fraction decompositions~\cite{Geyer:2015bja,Baadsgaard:2015hia,Baadsgaard:2015twa,Huang:2015cwh,Geyer:2017ela}, but it is usually extremely complicated to rewrite linear propagator integrands in terms of quadratic ones through a simple procedure since it involves reassembling terms of partial fractions and shifts of momenta, as well as including terms that vanish under integration. Although linear decompositions of loop amplitudes can be utilized directly using the Q-cut technology pioneered in Ref.~\cite{Baadsgaard:2015twa}, there can still be advantages associated with rewriting loop amplitude integrands, phrased in terms of linear propagators, as quadratic ones, to enjoy the computational technology developed over decades.\\[5pt]
A method for extending the ambitwistor string in $D=4$ at tree level to one-loop that give rise to integrands with quadratic propagators for supersymmetric theories was first proposed by Farrow and Lipstein in Ref. \cite{Farrow:2017eol}. Alternatively, scalar quadratic propagator integrands at loop level from linear propagator integrands can also be obtained using results of Refs.~\cite{Gomez:2016cqb,Gomez:2017lhy}. Here a double forward limit is employed with four extra loop legs instead of the usual two. When the two extra on-shell loop momenta are combined into two off-shell loop momenta, we gain a direct path to traditional quadratic integrands. In order to pioneer such a formalism for gauge theory integrands, we will draw on inspiration from Ref.~\cite{Gomez:2017lhy}, as well as the technology recently developed in Refs.~\cite{Bjerrum-Bohr:2016juj,Cardona:2016gon,Bjerrum-Bohr:2016axv,He:2016mzd,He:2017spx,Geyer:2017ela,Du:2017kpo}. \\[5pt]

As a reference for the integrands we generate from the procedure, we will refer to a $D=4-2\epsilon$ integral basis decomposition of one-loop amplitudes, discussed e.g. in~\cite{Melrose:1965kb,Bern:1993kr,Bern:1994zx,Elvang:2013cua}. Here amplitudes are decomposed into linear combinations of $n$-gon integrals $ I^{(l)}_n $ with quadratic propagators integrated over $D=4-2\epsilon$ dimensions and multiplied with coefficients $ C^{(l)}_n$,  
\begin{equation}\label{EqMasterIntegralExpansion}
A^{(1)}=\sum_{i} C^{(i)}_4 I_4^{(i)} +\sum_{j}C^{(j)}_{3}I^{(j)}_{3} +\sum_{k}C_2^{(k)}I^{(k)}_2+ \text{rational term}\,.
\end{equation}
The gauge-invariant integral coefficients $ C^{(l)}_n$ can be deduced from integrand reduction or four-dimensional unitarity, and we refer to, 
\begin{equation}\label{EqMasterIntegralExpansion2}
\tilde A^{(1)}=\sum_{i} C^{(i)}_4 I_4^{(i)} +\sum_{j}C^{(j)}_{3} I^{(j)}_{3} +\sum_{k}C_2^{(k)} I^{(k)}_2\,,
\end{equation}
as the four-dimensional cut-constructible part of the amplitude. The main focus of this paper will be the construction of a quadratic propagator integrand from which the four-dimensional cut-constructible part of the amplitude can be inferred, starting from a linear propagator integrand. \\[5pt]

The paper is organized as follows. In Sec.~\ref{YMSC} we outline how to obtain the representations for gauge theory integrands with linear propagators from the scattering equations. In Sec.~\ref{QP} we demonstrate how to use the results from Sec.~\ref{YMSC} to derive loop integrands with quadratic propagators and to validate the construction using four-dimensional unitarity.  Sec.~\ref{Sec.Discussion} contains our conclusions and discussions. There are three appendices. 

\section{Yang-Mills gauge theory linear propagator one-loop amplitudes from scattering equations}
\label{YMSC}
\begin{figure}[hbt!]
	\begin{tikzpicture}
	\begin{scope}[yshift=-1cm]
	\node { \includegraphics[scale=0.5]{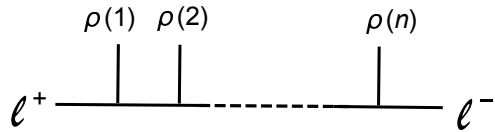} };
	\end{scope}
	\end{tikzpicture}\vskip-0.7cm
	\caption{{\small Half-ladder tree diagram associated with the one-loop color-kinematic numerator, $n[{\ell^+, \rho_n,\ell^-}]$ for off-shell momenta $\ell^+$ and $\ell^-$. }}
	\label{Linear-Quadratic0}\noindent
\end{figure}

Yang-Mills gauge theory tree amplitudes are computed in the scattering equation formalism from integrands of the type 
\begin{equation}
\mathcal{I}_n^{\rm (0)}(\alpha(1),\alpha(2),\ldots,\alpha(n))\;=\;{\rm PT}[\alpha(1),\alpha(2),\ldots,\alpha(n)] \times \mathcal{I}^{\rm Pf}_n\,,
\end{equation}
where
\begin{equation}\begin{split} \mathcal{I}^{\rm Pf}_n\;& \equiv\; \frac{(-1)^{i+j}}{z_{i}-z_{j}}\;\text{Pf}\Big[(\Psi)^{k_i k_j}_{k_i k_j}\Big]\,,
\end{split}\end{equation}
and $\text{Pf}\Big[(\Psi)^{k_i k_j}_{k_i k_j}\Big]$ denotes the Pfaffian of the
matrix $ \Psi=\left(\begin{array}{c|c}
A&-C^T\\ \hline\Tstrut\Bstrut
C&B
\end{array}\right)$ 
with rows and columns corresponding to legs $ k_i $ and $ k_j $ reduced and
\begin{equation}\begin{gathered}
\begin{aligned}\displaystyle
&A_{ij}\equiv\begin{cases}
\frac{2k_i\cdot k_j}{z_{i}-z_{j}}&
\\
0&
\end{cases}%
\ \ \ 
B_{ij}\equiv\begin{cases}
\frac{2\epsilon_i\cdot \epsilon_j\displaystyle}{z_{i}-z_{j}}&
\\
0&
\end{cases}
&C_{ij}\equiv\begin{cases}\frac{
	2\epsilon_i\cdot k_j}{z_{i}-z_{j}}&\quad\text{for }i\neq j\,,\\
-\sum_{l\neq i}\frac{2\epsilon_i\cdot k_l}{z_{i}-z_{l}}&\quad\text{for }i=j\,.
\end{cases}
\end{aligned}\end{gathered}\label{ABC_Def}
\end{equation}
Here $\epsilon_i$ denotes polarizations and we employ the short-hand notation $\sz_{ij} \equiv \sz_i - \sz_j$ as well as $k_{a_ 1a_2\cdots a_p} \equiv k_{a_1}+k_{a_2}+\cdots + k_{a_p}$. Given the ordering of the legs $\{\alpha(1),\alpha(2),\ldots,\alpha(n)\}\equiv \alpha(1,2,...,n)\equiv \alpha_n$ we define the Parke-Taylor factor
\begin{equation}\begin{split}
{\rm PT}[\alpha_n]\;&\equiv\; \frac{1}{(z_{\alpha(1)}-z_{\alpha(2)}) \, (z_{\alpha(2)}-z_{\alpha(3)}) \cdots (z_{\alpha({n-1})}-z_{\alpha(n)})\, (z_{\alpha(n)}-z_{\alpha(1)})}\,.
\end{split}\end{equation}
%
To use the scattering equation formalism to derive one-loop integrands in Yang-Mills theory, we will exploit knowledge about loop integrands in scalar $\phi^3$ theories in the forward limit. In Ref.~\cite{Baadsgaard:2015hia} it was shown how one-loop amplitudes with $\alpha_n$ external legs could be derived from Parke-Taylor integrands such as
\begin{equation}
\left(\sum_{\text{cyclic}}{\rm PT} [\ell^+,\alpha_n,\ell^-]\right)^2\,,
\end{equation}
where $\ell^+=-\ell^-\equiv \ell$ and where the sum is over cyclic permutations of $\alpha_n$. To extend this construction to Yang-Mills theory it is natural to use the following integrand
 \begin{equation}\begin{split}
 \!\!\!\!\mathcal{I}_n^{(1)}(\alpha_n)&=\frac{1}{\ell^2}\left(\sum_{\text{cyclic}}{\rm PT} [\ell^+,\alpha_n,\ell^-]\right)\times \left(\frac{(-1)^{n+2}}{\sz_{\ell^+}-\sz_{\ell^-}} \sum_{r=1}^{D-2}\ {\rm Pf}\! \left[ (\Psi_{n+2})^{{\ell^+}\, \ell^-}_{\ell^+\, \ell^-} \right]\right)\,.
\end{split}\end{equation}
Here $\Psi_{n+2}$ is a $2(n+2)\times 2(n+2)$ matrix with entries defined similarly to Eq. \eqref{ABC_Def}, supplemented with two additional opposite and equivalent off-shell loop leg rows and columns, $\ell^+$ and $\ell^-$. In the formula $D$ denotes the dimension of the space-time and the sum in $r$ runs over the physical polarization degrees of freedom of the off-shell legs.\\[5pt]
We have checked correspondence of the above integrand with the recent results of \cite{Geyer:2017ela,He:2017spx} and worked out several examples (up to four points) that demonstrate by direct computation that the application of the integration rules for scalar $\phi^3$ theories combined with the results of \cite{Bjerrum-Bohr:2016juj} allows identification of Yang-Mills one-loop amplitude results in the forward limit. \\[5pt]
A convenient way to expand a given integrand $\mathcal{I}_n^{(1)}(\alpha_n)$ in one-loop pure Yang-Mills theory is to exploit KLT orthogonality Refs.~\cite{BjerrumBohr:2010ta}
and to decompose the Pfaffian contribution in terms of products of Parke-Taylor factors and numerator coefficients that satisfy color-kinematics identities on the support of $(n+2)$ scattering equations Ref.~\cite{Cachazo:2013iea} (see also Refs.~\cite{Monteiro:2013rya,Du:2014uua,Du:2017kpo}),
\begin{equation}\label{truePF}
\frac{(-1)^{n+2}}{\sz_{\ell^+}-\sz_{\ell^-}} \, \sum_{r=1}^{D-2} \, {\rm Pf}\! \left[ (\Psi_{n+2})^{\ell^+\, \ell^-}_{\ell^+\, \ell^-} \right]
= \sum_{\rho_n\in S_n} n({\rho_n};\ell) \, \, {\rm PT}[\ell^+,\rho_n,\ell^-]\,.
\end{equation}
In the expression, we sum over permutations of $\rho_n$ and define
{
	\begin{equation}\label{ndef}
	n(\rho_n;\ell) \;\equiv\;
	\sum_{r=1}^{D-2}  n[{ \ell^+, \rho_n, \ell^- ]}\,,
	\end{equation}
}
\noindent
corresponding to the half-ladder tree diagram shown in Fig. \ref{Linear-Quadratic0}.\\[-4pt]
\vskip-0.2cm
\noindent In the sum over polarizations, it is useful to introduce two reference vectors $\eta$ and $q$ satisfying
 $q^2=\eps^r\cdot q=0$, $\eta\cdot \ell = \eta\cdot k_i= \eta\cdot \eps_i =0$ and $\eta^2=\ell^2$. From the completeness relation it then follows that $n{[ \ell^+, \rho_n, \ell^- ]}$ is invariant under shifts, $\ell \to \ell+\eta$ and that we can define
{
	\begin{equation}\label{loopshifting}
	\sum_{r=1}^{D-2} \eps_{\mu}^{r} \, (\eps_{\nu}^{r})^\dagger = g_{\mu\nu} - \frac{\tilde \ell_\mu \, q_\nu + \tilde\ell_\nu \, q_\mu }{\tilde\ell\cdot q}\equiv \Delta_{\mu\nu}\, , \qquad \tilde\ell = \ell+\eta,
	\end{equation}}
which allows the formulation of the compact rules
{
	\begin{equation}\label{CRlinear}
	\Delta_\mu^\mu = D-2, \quad \Delta_{\mu\nu} V^\mu W^\nu = V\cdot W, \quad \text{for any } V,W \in \left\{ k_i, \eps_i \right\}.
	\end{equation}}\noindent
%
An integrand at one-loop for pure Yang-Mills theory with linear propagators that utilizes color-kinematic identities is thus 
\begin{equation}
\mathcal{I}_n^{(1)}(\alpha_n) \; =\; \frac{1}{\ell^2}\, \times \; \sum_{\text{cyclic}}{\rm PT} [\ell^+,\alpha_n,\ell^-] \times \sum_{\rho_n\in S_n} \! n(\rho_n;\ell) \, \, {\rm PT}[\ell^+,\rho_n,\ell^-]. \qquad
\label{inspired}\end{equation}
It is important to note that color-kinematic numerators (also known as Bern-Carrasco-Johansson (BCJ) (for a review see Ref. \cite{Bern:2019prr}) numerators), do not have a unique representation (see for instance Ref.~\cite{Bern:2008qj,Mafra:2011kj,Fu:2017uzt}). We will in this presentation employ the numerator representation of Ref.~\cite{Fu:2017uzt}.
\section{Quadratic propagators}\label{QP}
\begin{figure}[hbt!]
\vskip-0.1cm\hskip0cm
	\begin{tikzpicture}
	\begin{scope}[xshift=7.9cm]
	\node { \includegraphics[scale=0.5]{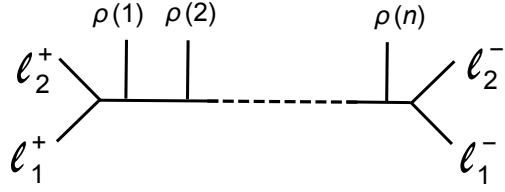} };
	\end{scope}
	\end{tikzpicture}\vskip-0.45cm
	\caption{{\small Half-ladder tree diagram associated with the one-loop color-kinematic numerator: $n[{\ell^+_1, \ell^+_2, \rho_n,\ell^-_1, \ell^-_2]}$ for on-shell $\ell_1^+$, $\ell_1^-$, $\ell_2^+$ and $\ell_2^-$.}} 
	\label{Linear-Quadratic}
\end{figure}

Armed with the machinery for integrands with linear propagators, we now develop a formalism for computing one-loop Yang-Mills integrands with traditional quadratic propagators. Since we work at the one-loop level it suffices to consider planar contributions. The basic idea is to consider a double forward limit of four massless on-shell gluons instead of the single forward limit we discussed so far. In the double forward limit we use $\ell\equiv\ell_1+\ell_2$ and $\ell_i^2=0$ with $\ell^2\neq 0$. Based on the above ideas and the linear propagator construction, we now propose the following integrand for Yang-Mills amplitudes in the double forward limit
\begin{eqnarray}\label{integrandCHY1L}
\mathcal{I}_{Q\, n}^{(1)}(\alpha_n) 
&\!=\!& {\cal PT}^{(1)}[\alpha_n]
  \sum_{\rho_n\in S_n} \!\! N(\rho_n;\ell_1,\ell_2) \, {\rm PT} [\ell_1^+\!,\ell_2^+\!,\rho_n,\ell_2^-\!,\ell_1^-]\,,
\qquad \qquad \\
{\cal PT}^{(1)}[\alpha_n] &\!\equiv\!& \sum_{\rm cyclic} \! {\rm PT} [\ell_2^+\!,\ell_1^+\!,\alpha_n,\ell_1^-\!,\ell_2^-] = \frac{\rm PT[\ell^+_1,\ell^+_2 , \ell^-_2, \ell^-_1]  } {\rm PT[\ell^+_1\!, \ell^-_1]} \sum_{\rm cyclic}  {\rm PT} [\ell_1^+\!,\alpha_n,\ell_1^-]\,,\qquad\qquad \ \ \\
N({\rho_n};\ell_1,\ell_2) &\!\equiv\!&\sum_{r_2=1}^{D-2}
\sum_{r_1=1}^{D-2}  n[{ \ell^+_1,\ell^+_2, \rho_n, \ell^-_2,\ell^-_1 ]}\, .
\qquad \qquad 
\end{eqnarray}
The sum in $r_1$ and $r_2$ run over the polarization degrees of freedom and $n[ \ell^+_1,\ell^+_2, $ $\rho_n, 
\ell^-_2,\ell^-_1 ]$, is the color-kinematic numerator corresponding to the propagator structure of the tree diagram in Fig. \ref{Linear-Quadratic}.
\vskip-0.0cm\noindent
We must in the double forward-limit at one-loop level specify additional rules for the sum over polarizations. 
They are as follows
\begin{eqnarray}\label{gluerelation}
\sum_{r_i=1}^{D-2} \eps_{\mu}^{r_i} \, (\eps_{\nu}^{r_i})^\dagger \equiv \Delta_{\mu\nu}^i, \quad i\in \{1,2\}\,,
\end{eqnarray} 
\vskip-0.1cm\noindent
thus the double-forward limit is derived from
\vskip-0.5cm
{
	\begin{eqnarray}
	(\Delta^i)_{\mu}^\mu&=&(\Delta^1)_{\mu\nu}(\Delta^2)^{\mu\nu}=D-2, \quad\\
	(\Delta^i)_{\mu\nu} V^\mu W^\nu &=& (\Delta^i)_{\mu}^{\a} (\Delta^j)_{\a\nu} V^\mu W^\nu = V\cdot W, \quad \nonumber\\
	(\Delta^i)_{\mu\nu} \ell_j^\mu V^\nu &=& V\cdot
	\ell_j ,
	\quad 
	(\Delta^i)_{\mu\nu} \ell_j^\mu \ell_j^\nu=
	(\Delta^i)_{\mu}^{\a} (\Delta^j)_{\a\nu} \ell_j^\mu V^\nu = 0
	\nonumber\\
	(\Delta^1)_{\mu}^{\a} (\Delta^2)_{\a\nu} \ell_2^\mu \ell_1^\nu
	&=&0, 
	\quad \text{for any } V,W \in \left\{ k_i, \eps_i \right\} \text{ and } i \neq j .
	\end{eqnarray}
}
\vskip-0.5cm\noindent
We now make the surprising observation that nontrivial algebraic connections exist between linear and quadratic numerators (although we do not have formal proof of this relation we have preformed extensive algebraic checks, up to $n=5$.)
\begin{equation}\label{NumeratorRel}
N(\rho_n;\ell_1, \ell_2)=\frac{\ell^2}{2}\, n(\rho_n;\ell) +(D-4)f(\ell,\tilde{\ell})\,.
\end{equation}
Here $\tilde \ell \equiv \ell_1-\ell_2$, and where $f(\ell,\tilde{\ell})$ is a nontrivial function. For instance, let us consider the simplest example, two particles, here we obtain,
\begin{align}\label{}
&
N(1,2;\ell_1, \ell_2)=\frac{\ell^2}{2}\, n(1,2;\ell) +(D-4)f_{12}(\ell,\tilde{\ell}), 
\end{align}
where (the numerator, $N(2,1;\ell_1, \ell_2)$, is obtained by relabeling, $1\leftrightarrow 2$.)
\begin{align}\label{}
&
n(1,2;\ell) = 
2 (D -2) \left[ 2 ( \epsilon_1 \cdot \ell) ( \epsilon_{2} \cdot \ell )  - (\epsilon_1\cdot \epsilon_2) (k_1 \cdot \ell) \right], 
\\
&
f_{1 2}(\ell,\tilde{\ell})= (\epsilon_1\cdot \epsilon_2) \left[ (k_1 \cdot \ell) - (k_1 \cdot \tilde\ell) \right]^2 .
\end{align}
If we are only interested in the parts of the integrand that can be derived from four-dimensional unitarity we do not have to consider the contribution $f(\ell,\tilde{\ell})$ and the linear and quadratic numerators become proportional. \\[5pt]
We thus arrive at the following simple prescription for the four-dimensional cut-constructible part of the Yang-Mills one-loop integrand 
\begin{eqnarray}\label{Prescription}
\!\!\!\!\!\!\!\!\!\!\!\! I^{(1)}(1,2,\ldots , n) &=& \int d\Omega \int \,\, d\mu_{(n+4)}\, \, \mathcal{I}_{Q\, n}^{(1)}(\alpha_n)\,,\\
 {\cal I}^{(1)}_{Q\, n}(\alpha_n)
&=& \frac{ \ell^2}{2}\times {\cal PT}^{(1)}[\alpha_n] \frac{ \text{PT}[\ell^+_1 ,\ell^+_2 , \ell^-_2, \ell^-_1]}{ \text{PT}[\ell^+_2, \ell^-_2]  }  \sum_{\rho_n\in S_n} n(\rho_n;\ell) \, {\rm PT}[\ell^+_2,\rho_n,\ell^-_2]\,,\nonumber
\end{eqnarray}
The identification: $\ell_i^+=-\ell_i^-$ and $\ell_1^+ + \ell_2^+=\ell$, is provided by
$\int \,d\Omega \equiv \int\, d^D(\ell_1^+ + \ell_2^+)\, \delta^{(D)}(\ell_1^+ + \ell_2^+ - \ell)\, d^{D} \ell_2^-\,\delta^{(D)} ( \ell_2^+ + \ell_2^- )\, d^{D} \ell_1^-\, \,\delta^{(D)} (\ell_1^+ + \ell_1^-)
$
and $\int d\mu_{(n+4)}$ integrate the $(n+4)$ scattering equations
\begin{eqnarray}
&&
{\cal S}_a = 
\sum_{i=1}^2
\left[
\frac{2\,k_a\cdot \ell_i^+}{\sz_{a \ell_i^+}} + \frac{2\, k_a\cdot \ell_i^-}{\sz_{a \ell_i^-}} \right]
+ \sum_{b=1 \atop b\neq a}^{n}  \frac{ 2\, k_a \cdot k_b }{\sz_{ ab}} =0, \quad a=1,...,n , \qquad\quad \nonumber \\
&&
{\cal S}_{\ell_{1}^\pm} = 
\frac{ 2\, \ell_{1}^{\pm} \cdot \ell_{1}^{\mp}}{\sz_{\ell_{1}^{\pm} \ell_{1}^{\mp}}} + \frac{ 2\, \ell_{1}^{\pm} \cdot \ell_{2}^+}{\sz_{ \ell_{1}^{\pm} \ell_{2}^+}}
+ \frac{ 2\, \ell_{1}^{\pm} \cdot \ell_{2}^-}{\sz_{\ell_{1}^{\pm} \ell_{2}^-}}
+ \sum_{b=1 }^{n}  \frac{2\, \ell_{1}^{\pm} \cdot k_b }{\sz_{ \ell_{1}^{\pm} b}} =0 \nonumber \\
&&
{\cal S}_{\ell_{2}^\pm} = 
\frac{ 2\, \ell_{2}^{\pm} \cdot \ell_{2}^{\mp}}{\sz_{\ell_{2}^{\pm} \ell_{2}^{\mp}}} + \frac{ 2\, \ell_{2}^{\pm} \cdot \ell_{1}^+}{\sz_{ \ell_{2}^{\pm} \ell_{1}^+}}
+ \frac{ 2\, \ell_{2}^{\pm} \cdot \ell_{1}^-}{\sz_{\ell_{2}^{\pm} \ell_{1}^-}}
+ \sum_{b=1 }^{n}  \frac{2\, \ell_{2}^{\pm} \cdot k_b }{\sz_{ \ell_{2}^{\pm} b}} 
=0.
\end{eqnarray}
To avoid any possible singular solutions of the scattering equations, we carry out the identification, $\ell_i^+=-\ell_i^-$ and $\ell_1^+ + \ell_2^+=\ell$, after integration over the scattering equations.\\[-20pt]
\section{Connection to four-dimensional unitarity cut}
To check the validity of the above quadratic Yang-Mills integrand we will now verify that it has to correct four-dimensional unitarity factorization into products of tree-level amplitudes when loop-propagators go on-shell. Without loss of generality, let us consider the double-cut (see {\it e.g.} Ref.~\cite{Bern:1994zx}) with branch-cut discontinuities at $\ell^2=0$ and $(\ell+k_1+k_2)^2=0$ as illustrated in Fig. 3.\\[15pt]
\begin{figure}[hbt!]
	\centering\vskip-0.3cm
	\begin{tikzpicture}
	\begin{scope}[]
	\node { \includegraphics[scale=0.33]{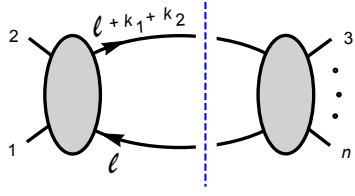} };
	\end{scope}
	\end{tikzpicture}\vskip-0.7cm
	\caption{{\small Double cut discontinuities: $\ell^2=0$ and $(\ell+k_1+k_2)^2=0$.}} 
	\label{UnitarityC}
\end{figure}\\[-5pt]
Inspired by Ref.~\cite{Cachazo:2015ksa}
we employ the coordinates
$
\sz_{\ell_1^\pm} = \sz_{\ell^\pm} + \tau\, \frac{\xi_{\ell^\pm}}{2} \, , \sz_{\ell_2^\pm} = \sz_{\ell^\pm} - \tau\, \frac{\xi_{\ell^\pm}}{2}\,
$
and integrate over the $ \xi_{\ell^\pm} $ part of the measure $d\mu_{(n+4)}$ in Eq.~\eqref{Prescription}.
 In the integration we sum over the residues corresponding to $(\ell_1^+\cdot \ell_2^+) \, \rightarrow \,0$ and $(\ell_1^-\cdot \ell_2^-) \, \rightarrow \,0$ for $ \tau\rightarrow \, 0\,.$
The result is
{
	\begin{eqnarray}\label{ellzero}
	\frac{1}{2\, \ell^2} \sum_{r=1}^{2} \! \int \!d\mu_{(n+2)}\, 
	\left(\sum_{\text{cyclic}}{\rm PT}[\ell^+,\alpha_n,\ell^-]\right)\,
	\frac{(-1)^{n+2}}{\sz_{\ell^+}-\sz_{\ell^-}} \, {\rm Pf}\! \left[ (\Psi_{n+2})^{\ell^+\, \ell^-}_{\ell^+\, \ell^-} \right]\,,
	\end{eqnarray}
}
\vskip-0.3cm\noindent
where $ \int \!d\mu_{(n+2)}$ denotes the residual integration over the scattering equations.
Now employing the technology of Refs.~\cite{Cachazo:2013hca,Bjerrum-Bohr:2018lpz,Gomez:2018cqg} we directly demonstrate that the integrand \eqref{Prescription} indeed yields the expected expression for the four-dimensional discontinuities of the one-loop amplitude. First, we note that by the integration, the result \eqref{ellzero} is a cyclic sum of on-shell Yang-Mills tree amplitudes.
The cut integral transforms as
$
\int d^4\ell\, \delta(\ell^2)\,\delta((\ell+k_1+k_2)^2) \, \, \rightarrow \, \, \int d^4\tilde\ell\, \delta(\tilde\ell^2)\,\delta((\tilde\ell-k_1-k_2)^2)
$
under the coordinate transformation $\ell \to \tilde \ell -k_1-k_2$. We now have identify the factorization channels corresponding to, $(\ell + k_1 + k_2)^2 = 2\, \ell\cdot k_{12} + s_{12}$ and $(\ell - k_1 - k_2)^2 = - 2\, \ell\cdot k_{12} + s_{12}$ in \eqref{ellzero}, as illustrated below,
\vspace{-0.1cm}
{
	\begin{eqnarray}\label{}
	\parbox[c]{9.9em}{\includegraphics[scale=0.27]{unitarityC.eps}} =
	\frac{1}{2}
	\left\{
	\parbox[c]{10.1em}{\includegraphics[scale=0.27]{unitarityC.eps}} +
	\parbox[c]{10.1em}{\includegraphics[scale=0.27]{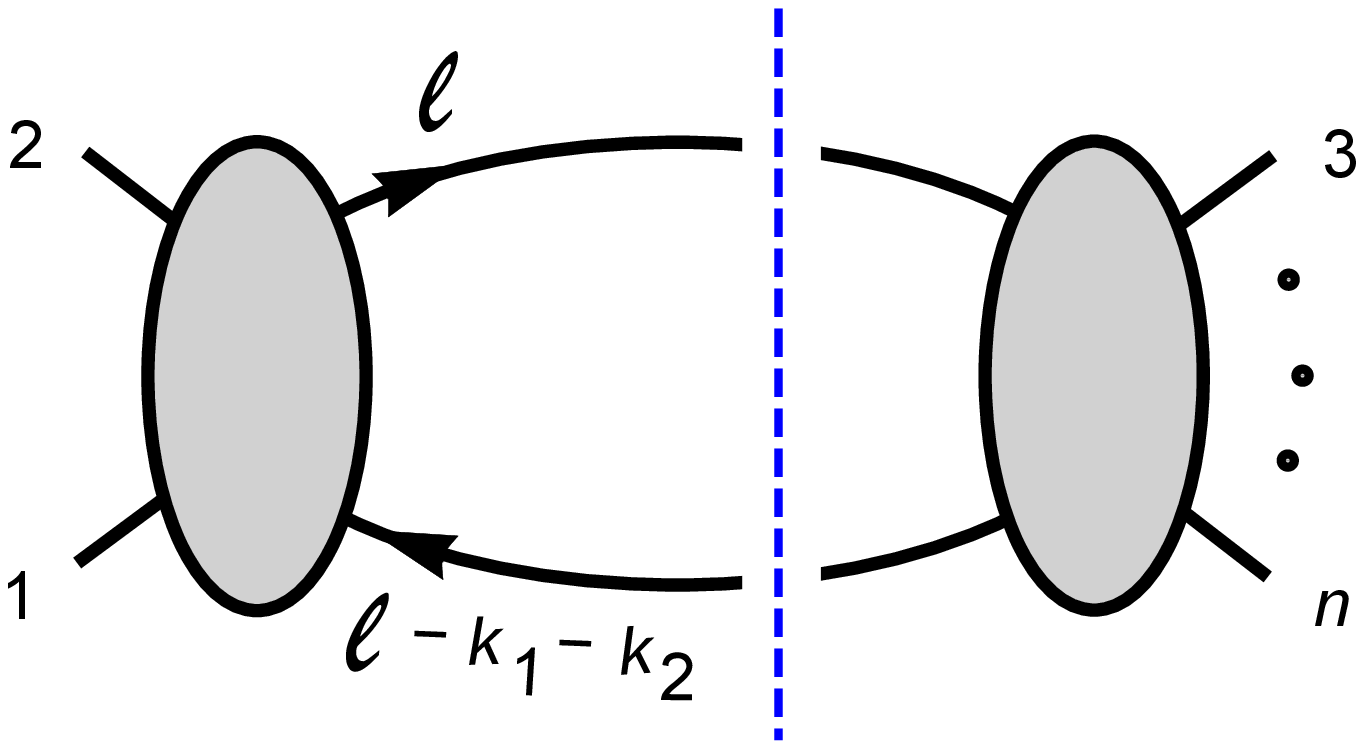}} \right\}.
	\nonumber
	\end{eqnarray}
}
\vskip-0.6cm\noindent
We arrive at
\!\!\!\!\vspace{-0.1cm}\hskip-.6cm\!\!\!\!\!\!\!\!\!\!
{
	\begin{eqnarray}\label{Ucutshift}
	\!\!\!\!\!\!\!\!
	&&\!\!\!\!\frac{1}{2} \sum_{r,s=1}^{2} \left[ \oint_{\Gamma} \frac{A^{(0)}(\ell^{-r},1,2, -(\ell+k_1+k_2)^{-s})\,\, A^{(0)}((\ell+k_1+k_2)^s,3,..., n, -\ell^r)}{\ell^2\,(\ell + k_1 + k_2)^2}
	\right.\\ \nonumber
	&&+
	\left. \oint_{\tilde\Gamma} \, \frac{A^{(0)}(\ell^{-r},3,\ldots,n, -(\ell-k_1-k_2)^{-s})\,\, A^{(0)}((\ell-k_1-k_2)^s,1,2, -\ell^r)}{\ell^2\,(\ell - k_1 - k_2)^2}
	\right]\,,\ \ \ \ \ \ \ \ \ 
	\end{eqnarray}
}
\vskip-0.4cm\noindent
where $\Gamma$ and $\tilde\Gamma$ are the contours circling the residues at, $\ell^2=0$ and $(\ell + k_1 + k_2)^2=0$ or $(\ell - k_1 - k_2)^2=0$ respectively. Finally by the shift of integration $\tilde\ell = \ell-k_1-k_2$, in the second integral, we land on the expected result
\vspace{-0.1cm}
{
	\begin{eqnarray}\label{}
	\!\!\!\!\!\!\!\!\!\!\!\!\!\!\! 
	\sum_{r,s=1}^{2} \oint_{\Gamma} \, \frac{A^{(0)}(\ell^{-r},1,2, -(\ell+k_1+k_2)^{-s})\,\, A^{(0)}((\ell+k_1+k_2)^s,3,..., n, -\ell^r)}{\ell^2\,(\ell + k_1 + k_2)^2}\,,
	\end{eqnarray}
}
\vskip-0.6cm\noindent
which validates our integrand construction \eqref{Prescription}.  
\section{Cut-constructible quadratic propagator integrand for Yang-Mills amplitudes from double forward limit}\label{subSec.Four-Point}
In this Sec., we compute the four-point integrand using the proposal in \eqref{Prescription}. As an explicit verification, we demonstrate 
exact agreement with the results previous obtained by Bern {\it et.al.} in \cite{Bern:2013yya}.
We start with \vskip-0.6cm
\begin{eqnarray}\label{}
{I}_4^{(1)}(1,2,3,4) = \int d\Omega \, \int d\mu_{(4+4)}\,\,  {\cal I}^{(1)}_{Q\, 4}(1,2,3,4)\,,
\end{eqnarray} \vskip-0.1cm\noindent
and after performing the integration, we immediately obtain the result
\begin{equation}\label{4point-btbb}
I_4^{(1)}(1,2,3,4)= \left( \, I_4+I_3+I_2 +I_1 \, \right) ,
\end{equation}
where the $I_4$, $I_3$, $I_2$ and $I_1$ are given by the expressions 
\begin{eqnarray}
I_4&=&  \sum_{{\rm cyclic}}
\frac{\frac{1}{2} \,n(1234;\ell)}{\ell^2(\ell+k_{1})^2(\ell+k_{12})^2(\ell+k_{123})^2} \,,\\
I_3&=& \frac{1}{2\,\ell^2}\sum_{{\rm cyclic}} \!
\!\frac{1}{s_{12}}\left[ \!\frac{n([1,2]34;\ell) }{(\ell\!+\!k_{12})^2(\ell\!+\!k_{123})^2}\!+\!\frac{n(4[1,2]3;\ell)}{(\ell\!+\!k_{4})^2(\ell\!+\!k_{412})^2}\!+\!\frac{n(34[1,2];\ell)}{(\ell\!+\!k_{3})^2(\ell\!+\!k_{34})^2}\right]\, ,\nonumber \\
I_2 &=& \nonumber
\frac{1}{2\,\ell^2}\left[ \frac{n([1,2][3,4];\ell)}{s_{12}\,s_{34}(\ell\!+\!k_{12})^2}\!+\!\frac{,n([3,4][1,2];\ell)}{s_{12}\,s_{34}(\ell\!+\!k_{34})^2} + \frac{n([2,3][4,1];\ell)}{s_{23}\,s_{41}(\ell\!+\!k_{23})^2}\!+\!\frac{n([4,1][2,3];\ell)}{s_{23}\,s_{41}(\ell\!+\!k_{41})^2} \right]\,,\nonumber\\
I_1&= &\frac{1}{2\,\ell^2}\!\sum_{{\rm cyclic}} 
\frac{1}{s_{234}}\left[ \frac{n(1[2,\![3,4]];\ell)}{s_{34}(\ell\!+\!k_{1})^2}\!+\!
\frac{n([2,\![3,4]]1;\ell)}{s_{34}(\ell\!+\!k_{234})^2}  
\!+\!
\frac{n(1[[2,3],4];\ell)}{s_{23}(\ell\!+\!k_{1})^2}	
+\frac{n([[2,3],4]1;\ell)}{s_{23}(\ell\!+\!k_{234})^2}\right] .\nonumber
\end{eqnarray}
We sum over cyclic permutations of the external legs $\{1,2,3,4\}$, and we have introduced the usual notation for the color-kinematic numerators
\begin{equation}\label{StandardBCJNotation}
\begin{aligned}
n([1,2]34;\ell)&=n(1234;\ell)-n(2134;\ell),\\
n([1,2][3,4];\ell)&=n(1234;\ell)-n(2134;\ell)+n(2143;\ell)-n(1243;\ell),\\
n(1[[2,3],4];\ell)&=n(1234;\ell)-n(1324;\ell)+n(1432;\ell)-n(1423;\ell), \\
n([[2,3],4]1;\ell)&=n(2341;\ell)-n(3241;\ell)+n(4321;\ell)-n(4231;\ell).
\end{aligned}
\end{equation}
To simplify the above expressions, we collect all equivalent diagrams by shifting the loop momenta. Thus, we arrive at the following one-loop integrands\vskip-0.2cm
\begin{eqnarray}\label{integrands}
I_4 &=&\frac{\,{\it{N}}_4(1234;\ell) }{\ell^2(\ell+k_1)^2(\ell+k_{12})^2(\ell+k_{123})^2}\,, \nonumber\\
I_3&=& \sum_{{\rm cyclic}} 
\frac{ {\it N}_3([1,2]34;\ell) }{s_{12}\,\ell^2\,(\ell+k_{12})^2(\ell+k_{123})^2}\, , \nonumber\\
I_2&=& 
\frac{ {\it N}_2([1,2][3,4];\ell)}{s_{12}\,s_{34}\,\ell^2(\ell+k_{12})^2} +
\frac{ {\it N}_2([2,3][4,1];\ell)}{s_{23}\,s_{41}\,\ell^2(\ell+k_{23})^2}\,, \nonumber\\
I_1&=& \sum_{{\rm cyclic}}
\left[ \frac{{\it N}_1(1[2,[3,4]];\ell)}{s_{234}\,s_{34}\,\ell^2(\ell+k_{1})^2}+
\frac{{\it N}_1(1[[2,3],4];\ell)}{s_{234}\,s_{23}\,\ell^2(\ell+k_{1})^2} 
\right]\,,
\end{eqnarray}\vskip-0.4cm\noindent
where\vskip-0.8cm
\begin{eqnarray}\label{eqNumeratorShifted}
{\it N}_4(1234;\ell)\ \ \ \ \ \ &=& \frac{1}{2} \left[ n(1234;\ell)\!+\!n(2341; \ell\!+\!k_1)\!+\!n(3412;\ell\!+\!k_{12})\!+\!n(4123;\ell\!+\!k_{123})\right]\,, \nonumber\\
{\it N}_3([1,2]34;\ell)\ \ \ &=& \frac{1}{2} \left[ n([1,2]34;\ell) +n(34[1,2];\ell+k_{12}) +n(4[1,2]3;\ell+k_{123}) \right]\,, \nonumber\\
		{\it N}_2([1,2][3,4];\ell)&=& \frac{1}{2}\, \left[ n([1,2][3,4];\ell) +n([3,4][1,2];\ell+k_{12}) \right]\,,\\
		{\it N}_1(1[2,[3,4]];\ell)&=& \frac{1}{2}\, \left[ n(1[2,[3,4]];\ell) + n([2,[3,4]]1 ;\ell+k_{1}) \right]\,, \nonumber\\
		{\it N}_1(1[[2,3],4];\ell)&=& \frac{1}{2}\, \left[ n(1[[2,3],4];\ell) + n([[2,3],4]1 ;\ell+k_{1}) \right]\,.\nonumber
\end{eqnarray}\vskip-0.1cm\noindent
We note that the leg-bubbles will vanish under integration using dimensional regularization. Clearly, 
${\it N}^{}_4(1234;\ell)-{\it N}^{}_4(2134;\ell) \neq {\it N}^{}_3([1,2]34;\ell),$
hence this is not a color-kinematic numerator representation. Nevertheless, this new representation has the advantage that all numerators are written in terms of one linear master numerator which is simple to perform computations on. The four-point master numerator is provided in the Appendix. Generalizations to higher multiplicities are expected but will not be pursued here. Since the cut-constructible part of the integrand is decomposable in a basis of box-, triangle-, and bubble type integrands, similar to the four-point case, we expect {\it a priori} the double forward limit for higher multiplicities to share certain generic features. The numerator relation Eq. \eqref{NumeratorRel} has been checked analytically till five points. \\[5pt]
\section{Quadruple and triple cut consistency of the results}\label{subSec.Four-Pointcut}
To check further the procedure and the consistency of the derived quadratic propagator integrand \eqref{integrands}, we will now consider the two types of generalized cuts illustrated in Fig. \ref{Qauduplec}, \ref{tripleC} (in Appendix \ref{CUTS}, we give more details on the computations). 
\begin{figure}[ht!!!!!]
	\centering
	\begin{tikzpicture}
	\begin{scope}[]
	\node { \includegraphics[scale=0.3]{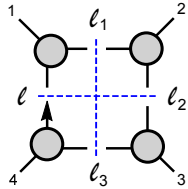} };
	\end{scope}
	\end{tikzpicture}
	\caption{{\small Quadruple cut given by the conditions, $\ell^2=(\ell_1)^2=(\ell_{2})^2=(\ell_3)^2=0$. We define, $\ell_1\equiv\ell+k_1, \, \ell_2\equiv \ell+k_{12}$ and $\ell_3\equiv\ell-k_4$.}} 
	\label{Qauduplec}
\end{figure}\\[-0pt]
The only nonvanishing helicity configurations in four dimensions are of the MHV-type, namely $ (--++) $ and $ (-+-+) $. For the helicity configuration $(1^-,2^-,3^+,4^+)$ we immediately identify the two quadruple cut contributions
\begin{equation}
d^{(1)}_{1^-,2^-,3^+,4^+}=d^{(2)}_{1^-,2^-,3^+,4^+}=i\,s_{12}\,s_{14}A^{(0)}(1^-,2^-,3^+,4^+)\,.
\end{equation}
In this case, there is no triple cut contribution. 

Now, we consider next the helicity configuration $ (1^-,2^+,3^-,4^+) $ where we again verify the quadruple cut solutions
\begin{equation}
\begin{aligned}
d^{(1)}_{1^-,2^+,3^-,4^+}&=i\,s_{12}\,s_{14}A^{(0)}(1^-,2^+,3^-,4^+),\\ d^{(2)}_{1^-,2^+,3^-,4^+}&=i\,s_{12}\,s_{14}\left(\frac{s_{12}^4+s_{14}^4}{s_{13}^4}\right)A^{(0)}(1^-,2^+,3^- ,4^+)\,.
\end{aligned}
\end{equation}
		 We solve the triple cut condition 
		\begin{equation}
		\ell^2=(\ell-k_4)^2=(\ell+k_1+k_2)^2=0\,,
		\end{equation}
		in the spinor helicity framework and identify the triangle contribution (see Fig. \ref{tripleC}).  
\begin{figure}[hbt!]
	\centering
	\begin{tikzpicture}
	\begin{scope}[]
	\node { \includegraphics[scale=0.3]{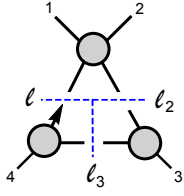} };
	\end{scope}
	\end{tikzpicture}
	\caption{{\small Triple cut given by, $\ell^2=(\ell_{2})^2=(\ell_3)^2=0$, where, $ \ell_2\equiv \ell+k_{12}$ and $\ell_3\equiv\ell-k_4$. }} 
	\label{tripleC}
\end{figure}\\[-0pt]
which is given by the expression
\begin{equation}
c_{\{1^-,2^+\},3^-,4^+} = -2is_{12}\frac{\braket{12}\braket{23}\braket{14}\braket{34}-2\braket{13}^2\braket{24}^2}{\braket{24}^4}.         
\end{equation}
All our results are in perfect agreement with the cut-constructible part of the integrand computed by using the one-loop numerator found by Bern {\it et.al.} in Ref.~\cite{Bern:2013yya} using the above box and triangle coefficients. 

An interesting point is the following. For the $ ([12]34) $ triangle, we find 
\begin{equation}
		\frac{ N_3([1,2]34;\ell) }{\ell^2 \left( \ell-k_4\right){}^2
			\left( \ell+k_1+k_2\right){}^2 s_{1 2}},
\end{equation}
with $ N_3([1,2]34;\ell)$ defined in \eqref{eqNumeratorShifted}. From this it appears that we na\"ively have a different number of box and triangle contributions and one could have expected a relative factor between box and triangle terms, since the triangle numerators are summed from three terms, whereas the box numerators are summed from four terms. However this is too simple, and to understand why, we have to consider the following feature of the obtained quadratic integrands. If we decompose quadratic box type integrands into linear box type integrands we have%
\vspace{-0.1cm}
{\small
\begin{eqnarray}\label{}
\!\!
\parbox[c]{6.2em}{\includegraphics[scale=0.3]{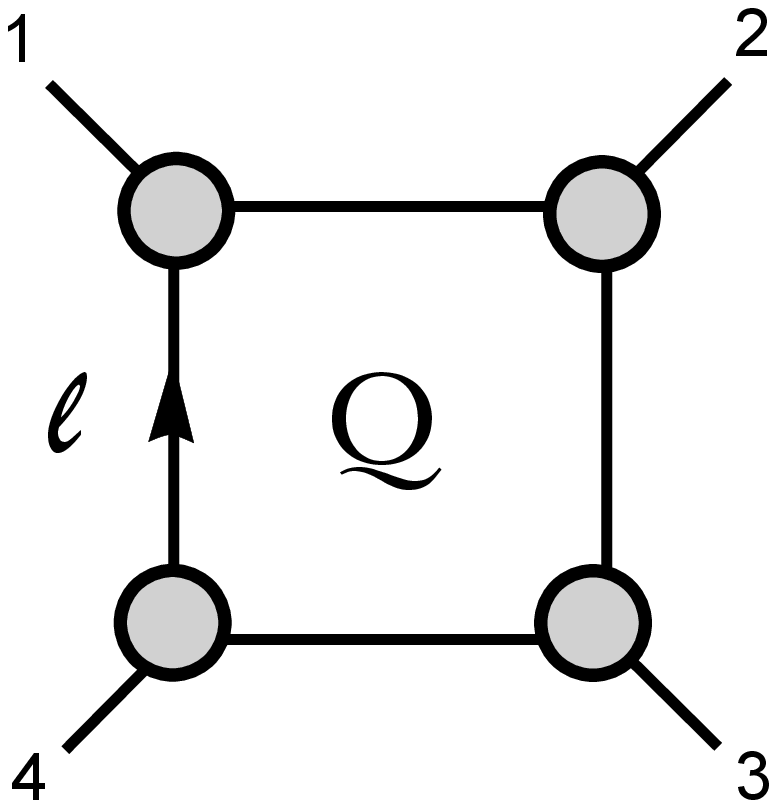}} 
\,
^{\underrightarrow{\quad \rm Partial\ Fraction\quad}}
\,\,
2\,
\left(\,\,\,\,
\parbox[c]{6.2em}{\includegraphics[scale=0.3]{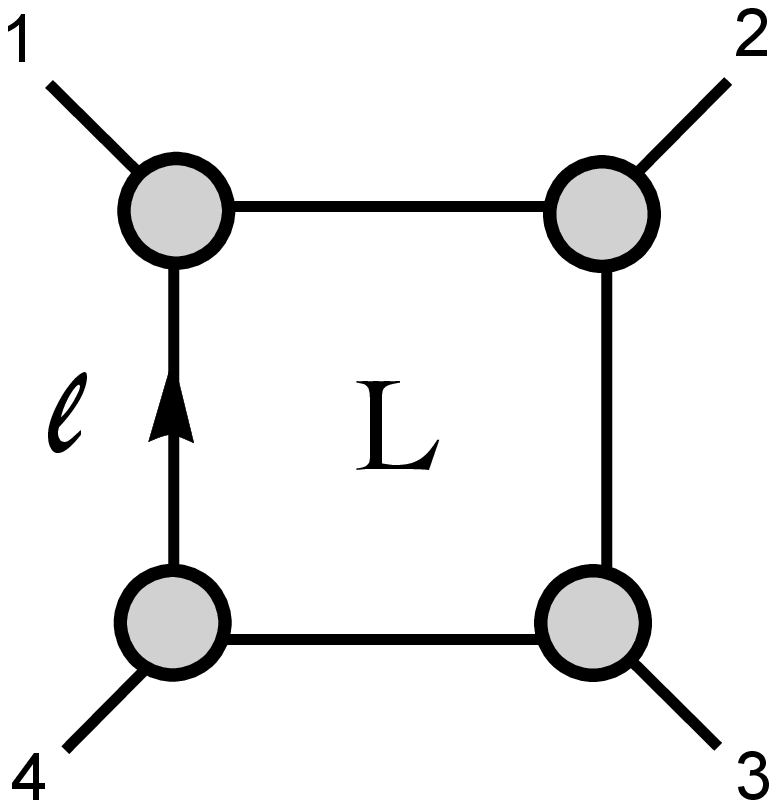}} 
\,\,\,\,
\right)
\,\,
+
\,\,
\,
\,\,
\parbox[c]{6.2em}{\includegraphics[scale=0.3]{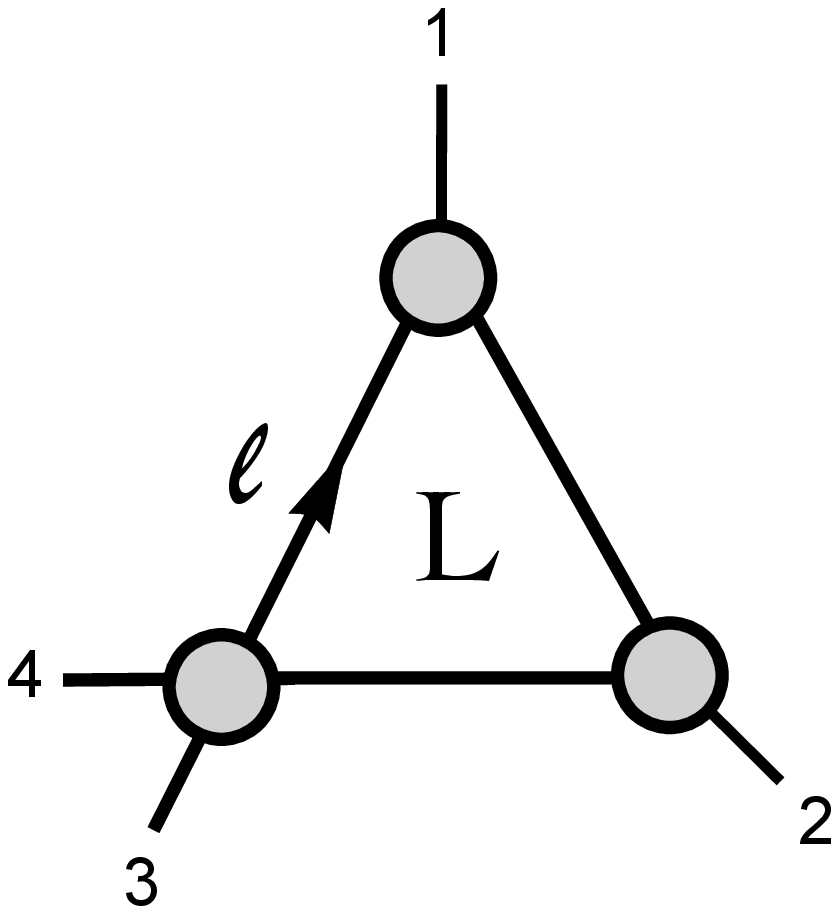}} 
\,\,\,\,
+ {\rm cyclic}\,.
\nonumber
\end{eqnarray}
}
\vskip-0.2cm\noindent
Meaning that the quadratic propagator box splits into the linear propagator box plus a contribution to the linear propagator triangle. By considering the difference between a quadratic box type integrand and a linear box type integrand we can qualify this statement. Considering the integrands for the case illustrated above one ends up with something proportional to 
\begin{equation}\label{qL}
		\frac{(\ell\cdot (k_1+k_2)+k_1\cdot k_2)+\ell \cdot k_4}{\ell^2 \ell\cdot k_1 \ell\cdot k_4 (\ell\cdot (k_1+k_2)+k_1\cdot k_2)}=\frac{1}{\ell^2 \ell\cdot k_1 \ell\cdot k_4 }+\frac{1}{\ell^2 \ell\cdot k_1 (\ell\cdot (k_1+k_2)+k_1\cdot k_2)}.
\end{equation} 
Of the two terms in $\eqref{qL}, $ one is homogenous in $ \ell $ and it, therefore, vanishes under integration (employing dimensional regularization) and but the other one will end up contributing as a triangle contribution in the integrand. This is validated by the triple cut. Quadratic propagator expressions for integrands from double forward limits and certain issues with such constructions have recently been the focus of Ref. \cite{Farrow:2020voh}. We would like to note, that for the four-point $D=4$ cut-integrand and the numerators, we consider here, the difference between the double and single forward limit boxes, is exactly a triangle as validated by the cuts and thus avoid any potential issues. Generalizations to integrands, where $f(\ell,\tilde{\ell})$ no longer can be neglected, has not been pursued in this paper.  
\section{Conclusion and discussion}\label{Sec.Discussion}
We have proposed a novel scattering equation construction, that we used to deduce the cut-constructible integrand part of one-loop Yang-Mills theories. The integrand we describe has quadratic propagators in the loop-momentum similar to a usual Feynman expansion due to the application of the double forward limit. To check our construction, we have verified the consistency of our integrand with four-dimensional unitarity. An interesting generalization of our formalism could be to investigate if it is also possible to capture the rational terms in Eq. \eqref{EqMasterIntegralExpansion} alike treatments found in Refs. e.g. \cite{NigelGlover:2008ur,Britto:2010um}. \\[5pt]  
It is possible to consider applications of our construction for integrands in the pure spinor formalism, see Ref.~\cite{Berkovits:2013xba} and well as for integrands from ambitwistor strings, see e.g. Refs.~\cite{He:2017spx,Mafra:2011kj,Geyer:2019hnn}. Another application could be for supersymmetric theories, where the cut-constructible integrand is sufficient to generate full amplitudes~\cite{Bern:1994zx}. An interesting idea is to extend the formalism considered here, to gravity amplitudes with massive sources, for instance, in the context of Ref.~\cite{Bjerrum-Bohr:2019nws}. Results for gravity loop amplitudes are increasingly becoming valuable input, for research in general relativity, and in such applications, only the cut-constructible parts of integrands are needed for the extraction of classical physics, see for instance Ref.~\cite{Bjerrum-Bohr:2018xdl}. We leave these ideas as potential directions for future research.

\subsection*{Acknowledgements}
We would like to thank Freddy Cachazo, Poul Henrik Damgaard, J.~A.~Farrow, Y.~Geyer, A.~E.~Lipstein and R.~Monteiro and R.~Stark-Much\~ao for useful discussions. We are also very grateful to Y. Geyer for providing us with her Mathematica package to compute color-kinematic numerators. This work was supported in part by the Danish National Research Foundation (DNRF91) as well as the Carlsberg Foundation. H.G. acknowledges partial support from University Santiago de Cali (USC). J.A acknowledges partial support by the  Villum Centre of  Excellence for the  Mathematics of  Quantum  Theory (QMATH).
\appendix

\section{Box and Triangle coefficients}\label{CUTS}
In this Appendix, we will provide some additional details on the box and triangle cut computations. 	
We obtained the coefficient of the box integral by evaluating the quadruple cut by at the solutions of the equations 
\begin{equation}
\ell^2=(\ell+k_1)^2=(\ell+k_1+k_2)^2=(\ell-k_4)^2=0\,,
\end{equation}
as shown in Fig. \ref{Qauduplec}. We have two cut solutions for the loop momentum. The helicity selection rules give us
\begin{align*}
	\ket{\ell^1}&\propto\ket{1}\propto\ket{\ell^1_1},
	\quad&\sket{\ell^1}\propto\sket{4}\propto\sket{\ell^1_3},\quad\text{and}\quad \ket{\ell^1_3}&\propto\ket{3}\propto\ket{\ell^1_2},\\
	\sket{\ell^2}&\propto\sket{1}\propto\sket{\ell^2_2},
	\quad&\ket{\ell^2}\propto\ket{4}\propto\ket{\ell^2_3},\quad\text{and}\quad \sket{\ell^2_3}&\propto\sket{3}\propto\sket{\ell^2_2},
	\end{align*}
	where the superscript denotes a particular solution to the cut and we have defined $ \ell_1=\ell+k_1 $ and $ \ell_3=\ell-k_4 $. From this we find the two solutions \begin{equation}
		\ket{\ell^1}\sbra{\ell^1}=\frac{\braket{34}}{\braket{31}}\ket{1}\sbra{4}\,,\qquad\text{and}\qquad\ket{\ell^2}\sbra{\ell^2}=\frac{\sbraket{34}}{\sbraket{31}}\ket{4}\sbra{1}\,,
	\end{equation}
	and we can then evaluate our result on the box cut. \\[5pt]
	For the triangle coefficient we focus on the nonvanishing helicity configuration $ (\{1^-,2^+\},3^-,$ $4^+) $ (all other configurations can be related to this case by relabeling.) We consider now the triple cut solution 
	\begin{equation}
	\ell^2=(\ell+k_1+k_2)^2=(\ell-k_4)^2=0\,,
	\end{equation}
such as it is illustrated in Fig. \ref{tripleC}.	
	
The triple cut has a leftover integration in four-dimensional unitarity, thus we parametrize the remaining integration by a parameter $t$. Because the triple cut has a quadratic constraint, there will be two solutions. We find  
\begin{equation}
	\ket{\ell_3^1}\propto\ket{3},\quad \sket{\ell_3^1}\propto\sket{4},\quad\text{and}\quad\ket{\ell_3^2}\propto\ket{4},\quad \sket{\ell_3^2}\propto\sket{3}\,,
	\end{equation}
where again the superscript denotes the particular solution to the cut. If we solve it up to the parameter $t$ we can write 
	\begin{equation}
	\ket{\ell_3^1}=t\ket{3},\quad \sket{\ell_3^1}=\sket{4},\quad\text{and}\quad\ket{\ell_3^2}=\ket{4},\quad\sket{\ell_3^2}=t\sket{3}.
	\end{equation}
	The rest of the loop-momentum spinors are then determined by momentum conservation. This solution exactly matches that of \cite{Forde:2007mi}. Computing the triangle coefficient, both the box-terms and triangle-terms in the amplitude will contribute. The contribution will be of the form 
	\begin{equation}
	\frac{{\it N}_4(1234;\ell) }{2k_1\cdot\ell},\quad\text{and}\quad \frac{ {\it N}_3([12]34;\ell)}{s_{12}}.
	\end{equation}
Evaluating these contributions on the cut, and extracting the residue at infinity we obtain the triangle coefficient\begin{equation}
c_{\{1^-,2^+\},3^-,4^+}=-2is_{12}\frac{\braket{12}\braket{23}\braket{14}\braket{34}-2\braket{13}^2\braket{24}^2}{\braket{24}^4}.
\end{equation} We have checked numerically that we have an exact match with the result of \cite{Bern:2013yya,Chester:2016ojq}.

\section{Four-point master numerator, $n(1234;\ell)$}\label{fourPnum}
To compute the color-kinematic numerators using the algorithm of Ref.~\cite{Fu:2017uzt}, it is necessary to choose a reference ordering. We specify the reference ordering, $ {\rm RO} $, in the following way by,
$
n(\rho_n |{\rm RO};\ell)$ and $N( \rho_n|{\rm RO} ;\ell_1,\ell_2).$
Symmetric combinations are \vskip-0.8cm
\begin{eqnarray}
&&
\!\!\!
n(\rho_n;\ell)
=\frac{1}{2\, (n!)}\sum_{ {\rm RO} \in S_n } \Big\{n(\rho_n|{\rm RO};\ell) 
+
(-1)^n n( \hat{\rho}_n |{\rm RO};-\ell)
\Big\},
\qquad \quad
\label{NumeratorDefLinear}
\\
&&
\!\!\!
N( \rho_n ;\ell_1,\ell_2)
=\frac{1}{2\, ( n! )}\sum_{ {\rm RO} \in S_n } \!\!\! \Big\{ N( \rho_n|{\rm RO};\ell_1,\ell_2)
+
(-1)^nN( \hat{\rho}_n|{\rm RO} ;-\ell_1,-\ell_2)
\Big\} ,\qquad\quad
\label{NumeratorDefQ}
\end{eqnarray}\vskip-0.2cm\noindent
where $ \hat{\rho}_n=\rho(n,n-1,...,1) $ is the reverse of $ \rho_n $ which appears in order to include the symmetry, $\ell \, \rightarrow \, -\ell$. The numerator $ n(1234;\ell) $ is computed in the symmetric combination.\\[5pt]
	\begin{equation}
	\begin{aligned}
	&n(1234;\ell)= \left[\vphantom{\frac{\frac{1}{2}}{9}}\left(2 \epsilon _1\cdot \epsilon _4 \epsilon _2\cdot \epsilon _3-\epsilon _1\cdot \epsilon _3 \epsilon
	_2\cdot \epsilon _4-3 \epsilon _1\cdot \epsilon _2 \epsilon _3\cdot \epsilon _4\right) \left(k_1\cdot
	k_2\right){}^2\right.\\&\left.+\left\{\frac{1}{4} \left(-5 k_1\cdot \ell+8 k_1\cdot k_4+5 k_4\cdot \ell\right) \epsilon _1\cdot \epsilon _4
	\epsilon _2\cdot \epsilon _3+\left[\vphantom{\frac{D}{2}}2 \epsilon _1\cdot \epsilon _4 \left(\epsilon _2\cdot k_1+\epsilon _2\cdots
	k_4\right)\right.\right.\right.\\&\left.\left.\left.+\left(5-\frac{D}{2}\right) \epsilon _1\cdot \ell \epsilon _2\cdot \epsilon _4\right] \epsilon _3\cdot \ell+\left(-2
	\epsilon _1\cdot \epsilon _4 \epsilon _2\cdot \ell-\frac{1}{6} \left((D-14) \epsilon _1\cdot \ell+6 \epsilon _1\cdot k_3\right)
	\epsilon _2\cdot \epsilon _4\right) \epsilon _3\cdot k_1\right.\right.\\&\left.\left.+\epsilon _2\cdot \epsilon _4 \left(\left(k_1\cdot k_3+k_3\cdot
	l\right) \epsilon _1\cdot \epsilon _3-2 \epsilon _1\cdot k_4 \epsilon _3\cdot \ell+\frac{1}{3} \left((D-14) \epsilon _1\cdot
	\ell+6 \epsilon _1\cdot k_4\right) \epsilon _3\cdot k_4\right)\right.\right.\\&\left.\left.-\frac{1}{6} \left[\vphantom{\frac{1}{2}}\epsilon _1\cdot \ell \left(-3 (D-10) \epsilon
	_2\cdot \ell-(D-14) \left(\epsilon _2\cdot k_1-2 \epsilon _2\cdot k_4\right)\right)\right.\right.\right.\\&\left.\left.\left.+6 \left(\epsilon _1\cdot k_3 \epsilon
	_2\cdot k_1+\epsilon _1\cdot k_4 \left(\epsilon _2\cdot k_1+2 \epsilon _2\cdot k_4\right)\right)\vphantom{\frac{1}{2}}\right] \epsilon _3\cdot
	\epsilon _4\right.\right.\\&\left.\left.+\frac{1}{6} \left[\vphantom{\frac{1}{2}}-\left(\epsilon _1\cdot \epsilon _3\right) \left(3 (D-2) \epsilon _2\cdot \ell+2 (D-2) \epsilon
	_2\cdot k_1-(D-14) \epsilon _2\cdot k_4\right)\right.\right.\right.\\&\left.\left.\left.-\left(6 (D-6) \epsilon _1\cdot \ell-2 \epsilon _1\cdot k_4+D \left(2 \epsilon
	_1\cdot k_3+\epsilon _1\cdot k_4\right)\right) \epsilon _2\cdot \epsilon _3+2 \epsilon _1\cdot \epsilon _2 \left[\vphantom{\sum}3 (D-6)
	\epsilon _3\cdot \ell\right.\right.\right.\right.\\&\left.\left.\left.\left.+(D-8) \left(\epsilon _3\cdot k_1-2 \epsilon _3\cdot k_4\right)\vphantom{\sum}\right]\vphantom{\frac{1}{2}}\right] \epsilon _4\cdot
	\ell+\left(\vphantom{\frac{1}{2}}\epsilon _1\cdot \epsilon _3 \left(2 \left(\epsilon _2\cdot \ell+\epsilon _2\cdot k_1\right)+\epsilon _2\cdot
	k_4\right)\right.\right.\right.\\&\left.\left.\left.-\frac{1}{6} (D+10) \epsilon _1\cdot \ell \epsilon _2\cdot \epsilon _3-2 \epsilon _1\cdot \epsilon _2 \left(\epsilon
	_3\cdot \ell+\epsilon _3\cdot k_1-\epsilon _3\cdot k_4\right)\right) \epsilon _4\cdot k_1\right.\right.\\&\left.\left.+\left(\epsilon _1\cdot \epsilon _3
	\epsilon _2\cdot k_4-\frac{1}{3} (D-2) \epsilon _1\cdot \ell \epsilon _2\cdot \epsilon _3+2 \epsilon _1\cdot \epsilon _2
	\epsilon _3\cdot k_4\right) \epsilon _4\cdot k_3\right\} k_1\cdot k_2\right.\\&\left.-\frac{1}{4} (D-2) \left(\left(\epsilon _1\cdot
	\epsilon _4 \epsilon _2\cdot \epsilon _3+\epsilon _1\cdot \epsilon _3 \epsilon _2\cdot \epsilon _4\right) \left(k_1\cdot
	\ell\right){}^2+\left(k_4\cdot \ell\right){}^2 \epsilon _1\cdot \epsilon _2 \epsilon _3\cdot \epsilon
	_4\right)\right.\\&\left.+\left[\left(-\left(k_4\cdot \ell\right) \epsilon _1\cdot \ell-\frac{2}{3} \left(-\left(k_1\cdot \ell\right)+2 k_1\cdot
	k_2+3 k_1\cdot k_3+k_3\cdot \ell\right) \epsilon _1\cdot k_3\right) \epsilon _2\cdot \epsilon _3\right.\right.\\&\left.\left.+\left(2 (D-2) \epsilon
	_1\cdot \ell \left(\epsilon _2\cdot \ell+\epsilon _2\cdot k_1\right)-4 \left(\epsilon _1\cdot k_3+\epsilon _1\cdot k_4\right)
	\epsilon _2\cdot k_1\right) \epsilon _3\cdot \ell\right.\right.\\
	&\left.\left.+\left(-2 \epsilon _1\cdot k_3 \left(\epsilon _2\cdot k_4-2 \epsilon _2\cdot
	\ell\right)-2 \left(\epsilon _1\cdot k_4-2 \epsilon _1\cdot \ell\right) \left(\epsilon _2\cdot k_1+\epsilon _2\cdot
	k_4\right)\right) \epsilon _3\cdot k_1\right.\right.\\&\left.\left.+\left[k_1\cdot k_4 \epsilon _1\cdot \epsilon _2+\left(-2 (D-4) \epsilon _1\cdot \ell+2
	\epsilon _1\cdot k_3+4 \epsilon _1\cdot k_4\right) \epsilon _2\cdot k_1\right.\right.\right.\\&\left.\left.\left.+\epsilon _1\cdot \ell \left(4 \epsilon _2\cdot k_4-2
	(D-2) \epsilon _2\cdot \ell\right)\right] \epsilon _3\cdot k_4\vphantom{\frac{1}{2}}\right] \epsilon _4\cdot \ell+k_1\cdot k_3 \left\{\vphantom{\sum}-\left(k_1\cdot
	l\right) \epsilon _1\cdot \epsilon _4 \epsilon _2\cdot \epsilon _3\right.\right.\\&\left.\left.+k_4\cdot \ell \epsilon _1\cdot \epsilon _4 \epsilon _2\cdot
	\epsilon _3-k_3\cdot \ell \epsilon _1\cdot \epsilon _3 \epsilon _2\cdot \epsilon _4+\left(2 \epsilon _1\cdot \epsilon _4
	\epsilon _2\cdot \ell-2 \epsilon _1\cdot \ell \epsilon _2\cdot \epsilon _4\right) \epsilon _3\cdot k_4\right.\right.\\&\left.\left.+\left(-\left(k_1\cdot
	k_2+2 k_1\cdot k_4\right) \epsilon _1\cdot \epsilon _2-2 \epsilon _1\cdot k_4 \epsilon _2\cdot \ell+2 \epsilon _1\cdot \ell
	\epsilon _2\cdot k_4\right) \epsilon _3\cdot \epsilon _4\right.\right.
	\end{aligned}
	\end{equation}
	\begin{equation*}
	\begin{aligned}
	&\left.\left.-\frac{1}{2} \left[8 \epsilon _1\cdot \epsilon _3 \epsilon _2\cdot
	\ell+4 \epsilon _1\cdot \epsilon _3 \epsilon _2\cdot k_1-8 \epsilon _1\cdot \ell \epsilon _2\cdot \epsilon _3+4 \epsilon _1\cdot
	k_4 \epsilon _2\cdot \epsilon _3-D \epsilon _1\cdot \epsilon _2 \epsilon _3\cdot \ell\right.\right.\right.\\&\left.\left.\left.+2 \epsilon _1\cdot \epsilon _2 \epsilon
	_3\cdot \ell+D \epsilon _1\cdot \epsilon _2 \epsilon _3\cdot k_4-6 \epsilon _1\cdot \epsilon _2 \epsilon _3\cdot k_4\right]
	\epsilon _4\cdot \ell\vphantom{\sum}\right\}+\left[\vphantom{\frac{1}{2}}\left(\left(k_1\cdot \ell+k_4\cdot \ell\right) \epsilon _1\cdot k_4\right.\right.\right.\\&\left.\left.\left.+2 \left(k_1\cdot k_2 \epsilon
	_1\cdot k_3+k_3\cdot \ell \epsilon _1\cdot k_4\right)\right) \epsilon _2\cdot \epsilon _3+\left(2 \epsilon _1\cdot k_4 \left(2
	\epsilon _2\cdot \ell+\epsilon _2\cdot k_1\right)\right.\right.\right.\\&\left.\left.\left.+2 \left(-2 \epsilon _1\cdot \ell+\epsilon _1\cdot k_3+\epsilon _1\cdot
	k_4\right) \epsilon _2\cdot k_4\right) \epsilon _3\cdot \ell+\left(2 k_1\cdot k_3 \epsilon _1\cdot \epsilon _2-2
	\left(\epsilon _1\cdot k_3+2 \epsilon _1\cdot k_4\right) \epsilon _2\cdot \ell\right.\right.\right.\\&\left.\left.\left.-2 \epsilon _1\cdot \ell \left(\epsilon _2\cdot
	k_1-\epsilon _2\cdot k_4\right)\right) \epsilon _3\cdot k_4\vphantom{\frac{1}{2}}\right] \epsilon _4\cdot k_1+\left[\vphantom{\frac{1}{2}}\frac{1}{2} \left(k_1\cdot
	l-k_4\cdot \ell\right) \epsilon _1\cdot k_3 \epsilon _2\cdot \epsilon _3\right.\right.\\&\left.\left.-2 \left(\left(k_1\cdot k_4 \epsilon _1\cdot
	k_3+k_1\cdot k_2 \epsilon _1\cdot k_4\right) \epsilon _2\cdot \epsilon _3+\epsilon _1\cdot k_4 \epsilon _2\cdot k_1
	\epsilon _3\cdot \ell\right)\right.\right.\\&\left.\left.+\left(2 \epsilon _1\cdot k_4 \epsilon _2\cdot \ell-2 \epsilon _1\cdot \ell \epsilon _2\cdot k_4\right)
	\epsilon _3\cdot k_1+k_1\cdot k_3 \epsilon _1\cdot \epsilon _2 \epsilon _3\cdot k_4\right.\right.\\&\left.\left.+\epsilon _1\cdot \ell \left(-\frac{2}{3}
	k_3\cdot \ell \epsilon _2\cdot \epsilon _3-4 \epsilon _2\cdot k_4 \epsilon _3\cdot \ell+2 \left(2 \epsilon _2\cdot \ell+\epsilon
	_2\cdot k_1\right) \epsilon _3\cdot k_4\right)\vphantom{\frac{1}{2}}\right] \epsilon _4\cdot k_3\right.\\&\left.+\frac{1}{24} k_1\cdot \ell \left\{\vphantom{\frac{1}{2}}12 \left(k_3\cdot
	\ell+k_4\cdot \ell\right) \epsilon _1\cdot \epsilon _4 \epsilon _2\cdot \epsilon _3-24 \left(k_1\cdot k_3+k_1\cdot k_4\right)
	\epsilon _1\cdot \epsilon _3 \epsilon _2\cdot \epsilon _4\right.\right.\\&\left.\left.-4 (D-2) \left(3 \epsilon _1\cdot \epsilon _4 \epsilon _2\cdot \ell-3
	\epsilon _1\cdot \ell \epsilon _2\cdot \epsilon _4+2 \left(\epsilon _1\cdot \epsilon _4 \epsilon _2\cdot k_1+\epsilon _1\cdot
	k_3 \epsilon _2\cdot \epsilon _4+\epsilon _1\cdot k_4 \epsilon _2\cdot \epsilon _4\right)\right) \epsilon _3\cdot \ell\right.\right.\\&\left.\left.+4
	\left((D-2) \epsilon _1\cdot \epsilon _4 \epsilon _2\cdot \ell-6 \epsilon _1\cdot \epsilon _4 \left(\epsilon _2\cdot
	k_1+\epsilon _2\cdot k_4\right)+(D-2) \epsilon _1\cdot \ell \epsilon _2\cdot \epsilon _4+6 \epsilon _1\cdot k_3 \epsilon
	_2\cdot \epsilon _4\right) \epsilon _3\cdot k_1\right.\right.\\&\left.\left.+2 \left(\vphantom{\sum}6 (D-2) \epsilon _1\cdot \epsilon _4 \epsilon _2\cdot \ell+3 \epsilon
	_1\cdot \epsilon _4 \left((D-6) \epsilon _2\cdot k_1-4 \epsilon _2\cdot k_4\right)-4 (D-2) \epsilon _1\cdot \ell \epsilon
	_2\cdot \epsilon _4\right.\right.\right.\\&\left.\left.\left.+3 (D-2) \left(\epsilon _1\cdot k_3+\epsilon _1\cdot k_4\right) \epsilon _2\cdot \epsilon _4\vphantom{\sum}\right)
	\epsilon _3\cdot k_4+2 (D-2) \left(\vphantom{\sum}4 \epsilon _1\cdot k_3 \epsilon _2\cdot \ell+4 \epsilon _1\cdot \ell \left(\epsilon _2\cdot
	k_1+\epsilon _2\cdot k_4\right)\right.\right.\right.\\&\left.\left.\left.-3 \left(k_1\cdot k_3 \epsilon _1\cdot \epsilon _2+\epsilon _1\cdot k_3 \epsilon _2\cdot
	k_4+\epsilon _1\cdot k_4 \left(\epsilon _2\cdot k_1+\epsilon _2\cdot k_4\right)\right)\vphantom{\sum}\right) \epsilon _3\cdot \epsilon
	_4\right.\right.\\&\left.\left.+3 k_1\cdot k_2 \left(-10 \epsilon _1\cdot \epsilon _3 \epsilon _2\cdot \epsilon _4+22 \epsilon _1\cdot \epsilon _2
	\epsilon _3\cdot \epsilon _4+D \left(\epsilon _1\cdot \epsilon _4 \epsilon _2\cdot \epsilon _3+\epsilon _1\cdot \epsilon _3
	\epsilon _2\cdot \epsilon _4-3 \epsilon _1\cdot \epsilon _2 \epsilon _3\cdot \epsilon _4\right)\right)\right.\right.\\&\left.\left.+4 \left(-(D-2)
	\epsilon _1\cdot \epsilon _3 \left(3 \epsilon _2\cdot \ell+2 \epsilon _2\cdot k_1\right)-\left[-3 (D-2) \epsilon _1\cdot \ell-4
	\epsilon _1\cdot k_4\right.\right.\right.\right.\\&\left.\left.\left.\left.+2 D \left(\epsilon _1\cdot k_3+\epsilon _1\cdot k_4\right)\right] \epsilon _2\cdot \epsilon _3+(D-2)
	\epsilon _1\cdot \epsilon _2 \left(-6 \epsilon _3\cdot \ell+2 \epsilon _3\cdot k_1+7 \epsilon _3\cdot k_4\right)\right)
	\epsilon _4\cdot \ell\right.\right.\\&\left.\left.+2 \left[2 \epsilon _1\cdot \epsilon _3 \left((D-2) \epsilon _2\cdot \ell+6 \epsilon _2\cdot k_4\right)+2
	(D-2) \epsilon _1\cdot \ell \epsilon _2\cdot \epsilon _3\right.\right.\right.\\&\left.\left.\left.-(D-2) \epsilon _1\cdot \epsilon _2 \left(3 \epsilon _3\cdot k_4-4
	\epsilon _3\cdot \ell\right)\right] \epsilon _4\cdot k_1+2 \left[\vphantom{\sum}-\left(\epsilon _1\cdot \epsilon _3\right) \left(2 (D-2)
	\epsilon _2\cdot \ell\right.\right.\right.\right.\\&\left.\left.\left.\left.+3 (D-2) \epsilon _2\cdot k_1-12 \epsilon _2\cdot k_4\right)-\left(-4 (D-2) \epsilon _1\cdot \ell-6 \epsilon
	_1\cdot k_4+3 D \left(\epsilon _1\cdot k_3+\epsilon _1\cdot k_4\right)\right) \epsilon _2\cdot \epsilon _3\right.\right.\right.\\&\left.\left.\left.+\epsilon _1\cdot
	\epsilon _2 \left(2 (D-2) \epsilon _3\cdot \ell+3 (D-2) \epsilon _3\cdot k_1-24 \epsilon _3\cdot k_4\right)\vphantom{\sum}\right] \epsilon
	_4\cdot k_3\vphantom{\frac{1}{2}}\right\}\right.\\&\left.
	+k_1\cdot k_4 \left(-\left(k_1\cdot \ell\right) \epsilon _1\cdot \epsilon _4 \epsilon _2\cdot \epsilon _3-2
	k_3\cdot \ell \epsilon _1\cdot \epsilon _4 \epsilon _2\cdot \epsilon _3-k_4\cdot \ell \epsilon _1\cdot \epsilon _4 \epsilon
	_2\cdot \epsilon _3\right.\right.\\&\left.\left.-k_1\cdot k_2 \epsilon _1\cdot \epsilon _3 \epsilon _2\cdot \epsilon _4+\left[-3 k_1\cdot k_2 \epsilon
	_1\cdot \epsilon _2+2 \epsilon _1\cdot k_3 \epsilon _2\cdot \ell-2 \epsilon _1\cdot k_4 \epsilon _2\cdot k_1-2 \epsilon
	_1\cdot k_3 \epsilon _2\cdot k_4\right.\right.\right.\\&\left.\left.\left.-2 \epsilon _1\cdot k_4 \epsilon _2\cdot k_4+2 \epsilon _1\cdot \ell \left(\epsilon _2\cdot
	k_1+\epsilon _2\cdot k_4\right)\right] \epsilon _3\cdot \epsilon _4+\epsilon _3\cdot \ell \left[\vphantom{\frac{1}{2}}-4 \epsilon _1\cdot \epsilon
	_4 \epsilon _2\cdot \ell-2 \epsilon _1\cdot \epsilon _4 \epsilon _2\cdot k_1\right.\right.\right.\\&\left.\left.\left.-2 \epsilon _1\cdot k_3 \epsilon _2\cdot \epsilon
	_4-2 \left(\epsilon _1\cdot k_4-2 \epsilon _1\cdot \ell\right) \epsilon _2\cdot \epsilon _4+\frac{1}{2} (D-2) \epsilon _1\cdot
	\epsilon _2 \epsilon _4\cdot \ell\right]\right.\right.
	\end{aligned}
	\end{equation*}
	\begin{equation*}
	\begin{aligned}
	&\left.\left.+\epsilon _3\cdot k_4 \left(\vphantom{\frac{1}{2}}4 \epsilon _1\cdot \epsilon _4 \epsilon _2\cdot \ell+2
	\epsilon _1\cdot \epsilon _4 \epsilon _2\cdot k_1+2 \epsilon _1\cdot k_3 \epsilon _2\cdot \epsilon _4+2 \left(\epsilon
	_1\cdot k_4-2 \epsilon _1\cdot \ell\right) \epsilon _2\cdot \epsilon _4\right.\right.\right.\\&\left.\left.\left.-\frac{1}{2} D \epsilon _1\cdot \epsilon _2 \epsilon
	_4\cdot \ell\right)+\left(-2 \epsilon _1\cdot \epsilon _3 \epsilon _2\cdot \ell-2 \epsilon _1\cdot \epsilon _3 \epsilon _2\cdot
	k_1+2 \epsilon _1\cdot \ell \epsilon _2\cdot \epsilon _3-2 \epsilon _1\cdot k_4 \epsilon _2\cdot \epsilon _3\right.\right.\right.\\&\left.\left.\left.+2 \epsilon
	_1\cdot \epsilon _2 \epsilon _3\cdot \ell+2 \epsilon _1\cdot \epsilon _2 \epsilon _3\cdot k_1+\epsilon _1\cdot \epsilon _2
	\epsilon _3\cdot k_4\right) \epsilon _4\cdot k_3\right)+\frac{1}{24} k_4\cdot \ell \left\{\vphantom{\frac{1}{2}}-48 k_1\cdot k_3 \epsilon _1\cdot
	\epsilon _3 \epsilon _2\cdot \epsilon _4\right.\right.\\&\left.\left.+4 (D-2) \left(3 \epsilon _1\cdot \epsilon _4 \epsilon _2\cdot \ell+3 \epsilon _1\cdot
	\epsilon _4 \epsilon _2\cdot k_1+\epsilon _1\cdot \epsilon _4 \epsilon _2\cdot k_4+6 \epsilon _1\cdot \ell \epsilon _2\cdot
	\epsilon _4+\epsilon _1\cdot k_3 \epsilon _2\cdot \epsilon _4\right.\right.\right.\\&\left.\left.\left.+2 \epsilon _1\cdot k_4 \epsilon _2\cdot \epsilon _4\right)
	\epsilon _3\cdot \ell+24 \left(\epsilon _1\cdot \epsilon _4 \left(\epsilon _2\cdot k_1+\epsilon _2\cdot k_4\right)+\frac{1}{6}
	(D-2) \epsilon _1\cdot \ell \epsilon _2\cdot \epsilon _4\right.\right.\right.\\&\left.\left.\left.+2 \epsilon _1\cdot k_3 \epsilon _2\cdot \epsilon _4\vphantom{\frac{1}{2}}\right) \epsilon
	_3\cdot k_1-2 \left[4 (D-2) \epsilon _1\cdot \epsilon _4 \epsilon _2\cdot \ell+3 \epsilon _1\cdot \epsilon _4 \left((D-6)
	\epsilon _2\cdot k_1-4 \epsilon _2\cdot k_4\right)\right.\right.\right.\\&\left.\left.\left.+8 (D-2) \epsilon _1\cdot \ell \epsilon _2\cdot \epsilon _4+3 (D-2)
	\left(\epsilon _1\cdot k_3+\epsilon _1\cdot k_4\right) \epsilon _2\cdot \epsilon _4\right] \epsilon _3\cdot k_4\right.\right.\\&\left.\left.+\left[\vphantom{\sum}3
	(D-2) \left(3 k_1\cdot k_3+k_1\cdot k_4\right) \epsilon _1\cdot \epsilon _2-4 (D-2) \left(\epsilon _1\cdot k_3-\epsilon
	_1\cdot k_4\right) \epsilon _2\cdot \ell\right.\right.\right.\\&\left.\left.\left.+4 (D-2) \epsilon _1\cdot \ell \left(3 \epsilon _2\cdot \ell+6 \epsilon _2\cdot k_1+4
	\epsilon _2\cdot k_4\right)+6 \epsilon _1\cdot k_4 \left((D-6) \epsilon _2\cdot k_1+(D-2) \epsilon _2\cdot k_4\right)\right.\right.\right.\\&\left.\left.\left.+3
	\epsilon _1\cdot k_3 \left(2 (D-2) \epsilon _2\cdot k_4-8 \epsilon _2\cdot k_1\right)\vphantom{\sum}\right] \epsilon _3\cdot \epsilon _4-3
	k_1\cdot k_2 \left[-2 \epsilon _1\cdot \epsilon _3 \epsilon _2\cdot \epsilon _4+8 \epsilon _1\cdot \epsilon _2 \epsilon
	_3\cdot \epsilon _4\right.\right.\right.\\&\left.\left.\left.+D \left(\epsilon _1\cdot \epsilon _4 \epsilon _2\cdot \epsilon _3+\epsilon _1\cdot \epsilon _3 \epsilon
	_2\cdot \epsilon _4-4 \epsilon _1\cdot \epsilon _2 \epsilon _3\cdot \epsilon _4\right)\right]+6 k_3\cdot \ell \left[-2
	\epsilon _1\cdot \epsilon _3 \epsilon _2\cdot \epsilon _4\right.\right.\right.\\&\left.\left.\left.+2 \epsilon _1\cdot \epsilon _2 \epsilon _3\cdot \epsilon _4+D
	\left(\epsilon _1\cdot \epsilon _4 \epsilon _2\cdot \epsilon _3+\epsilon _1\cdot \epsilon _3 \epsilon _2\cdot \epsilon
	_4-\epsilon _1\cdot \epsilon _2 \epsilon _3\cdot \epsilon _4\right)\right]-6 k_1\cdot \ell \left[\vphantom{\sum}D \left(\epsilon _1\cdot
	\epsilon _4 \epsilon _2\cdot \epsilon _3\right.\right.\right.\right.\\&\left.\left.\left.\left.+\epsilon _1\cdot \epsilon _3 \epsilon _2\cdot \epsilon _4+\epsilon _1\cdot
	\epsilon _2 \epsilon _3\cdot \epsilon _4\right)-2 \left(\epsilon _1\cdot \epsilon _3 \epsilon _2\cdot \epsilon _4+\epsilon
	_1\cdot \epsilon _2 \epsilon _3\cdot \epsilon _4\right)\vphantom{\sum}\right]+12 \left(\vphantom{\sum}D \epsilon _1\cdot \ell \epsilon _2\cdot \epsilon
	_3\right.\right.\right.\\&\left.\left.\left.+(D-2) \epsilon _1\cdot \epsilon _2 \left(\epsilon _3\cdot k_4-\epsilon _3\cdot \ell\right)\vphantom{\sum}\right) \epsilon _4\cdot \ell+2
	(D-2) \left[2 \epsilon _1\cdot \ell \epsilon _2\cdot \epsilon _3\right.\right.\right.\\&\left.\left.\left.+\epsilon _1\cdot \epsilon _2 \left(3 \epsilon _3\cdot k_4-4
	\epsilon _3\cdot \ell\right)\right] \epsilon _4\cdot k_1+2 \left[\vphantom{\sum}(D-2) \epsilon _1\cdot \epsilon _3 \left(4 \epsilon _2\cdot
	\ell+3 \epsilon _2\cdot k_1\right)\right.\right.\right.\\&\left.\left.\left.+\left(8 (D-2) \epsilon _1\cdot \ell-6 \epsilon _1\cdot k_4+3 D \left(\epsilon _1\cdot
	k_3+\epsilon _1\cdot k_4\right)\right) \epsilon _2\cdot \epsilon _3-\epsilon _1\cdot \epsilon _2 \left(4 (D-2) \epsilon
	_3\cdot \ell\right.\right.\right.\right.\\&\left.\left.\left.\left.+3 (D-2) \epsilon _3\cdot k_1+12 \epsilon _3\cdot k_4\right)\vphantom{\sum}\right] \epsilon _4\cdot k_3\vphantom{\frac{1}{2}}\right\}+\frac{1}{12}
	k_3\cdot \ell \left\{\vphantom{\frac{1}{2}}-6 k_4\cdot \ell \epsilon _1\cdot \epsilon _4 \epsilon _2\cdot \epsilon _3\right.\right.\\&\left.\left.+2 \epsilon _2\cdot \epsilon _4
	\left(6 \left(k_1\cdot k_4 \epsilon _1\cdot \epsilon _3+\epsilon _1\cdot k_3 \epsilon _3\cdot k_1\right)+(D-2) \epsilon
	_1\cdot \ell \left(3 \epsilon _3\cdot \ell+\epsilon _3\cdot k_1-2 \epsilon _3\cdot k_4\right)\right)\right.\right.\\&\left.\left.+2 \left[6 \left(2 k_1\cdot
	k_2 \epsilon _1\cdot \epsilon _2+\epsilon _1\cdot k_3 \epsilon _2\cdot k_1+\epsilon _1\cdot k_4 \epsilon _2\cdot
	k_1\right)-(D-2) \epsilon _1\cdot \ell \left(3 \epsilon _2\cdot \ell+\epsilon _2\cdot k_1\right.\right.\right.\right.\\&\left.\left.\left.\left.-2 \epsilon _2\cdot k_4\right)\right]
	\epsilon _3\cdot \epsilon _4-3 k_1\cdot \ell \left[-2 \epsilon _1\cdot \epsilon _3 \epsilon _2\cdot \epsilon _4+2 \epsilon
	_1\cdot \epsilon _2 \epsilon _3\cdot \epsilon _4+D \left(\epsilon _1\cdot \epsilon _4 \epsilon _2\cdot \epsilon _3\right.\right.\right.\right.\\&\left.\left.\left.\left.+\epsilon
	_1\cdot \epsilon _3 \epsilon _2\cdot \epsilon _4-\epsilon _1\cdot \epsilon _2 \epsilon _3\cdot \epsilon _4\right)\right]+2
	\left[(D-2) \epsilon _1\cdot \epsilon _3 \left(3 \epsilon _2\cdot \ell+2 \epsilon _2\cdot k_1-\epsilon _2\cdot
	k_4\right)\right.\right.\right.\\&\left.\left.\left.+\left(6 (D-2) \epsilon _1\cdot \ell-2 \epsilon _1\cdot k_4+D \left(2 \epsilon _1\cdot k_3+\epsilon _1\cdot
	k_4\right)\right) \epsilon _2\cdot \epsilon _3-(D-2) \epsilon _1\cdot \epsilon _2 \left(3 \epsilon _3\cdot \ell\right.\right.\right.\right.\\&\left.\left.\left.\left.+2 \epsilon
	_3\cdot k_1-\epsilon _3\cdot k_4\right)\right] \epsilon _4\cdot \ell+2 \left((D-2) \epsilon _1\cdot \ell \epsilon _2\cdot
	\epsilon _3-6 \epsilon _1\cdot \epsilon _3 \epsilon _2\cdot k_4\right) \epsilon _4\cdot k_1\right.\right.\\&\left.\left.+4 \left(D \epsilon _1\cdot \ell
	\epsilon _2\cdot \epsilon _3-3 \left(\epsilon _1\cdot \epsilon _3 \epsilon _2\cdot k_4+\epsilon _1\cdot \epsilon _2
	\epsilon _3\cdot k_4\right)\right) \epsilon _4\cdot k_3\vphantom{\frac{1}{2}}\right\}\vphantom{\frac{\frac{1}{2}}{9}}\right]
	\end{aligned}
	\end{equation*}

\section{Comments on generalizations for higher loop amplitudes}\label{Sec.Generalizations}
The one-loop double forward-limit was originally obtained by considering two-loop amplitudes in the ambitwistor string theory. Considering correlation functions over a genus two Riemann surface, one may localize on the boundary of the moduli space by applying the global residue theorem. Mathematically, we are thus considering the hyperelliptic curve
\begin{equation}
y^2=(z-a_1)(z-a_2)(z-\l_1)(z-\l_2)(z-\l_3),
\end{equation}
where $a_1\neq a_2$ are two fixed branch points and $(\l_1,\l_2,\l_3)$ parametrize the moduli space, 
and we focus at the degeneration points, $\l_1 = a_1$
and $\l_2 = a_2$ pinching the ${\rm A}$-cycles. It should be noted that many of the other singularities cancel out after computing the CHY integrals. The two global holomorphic forms on this curve, 
%
\begin{equation}
\Omega_1 \, dz = \frac{dz}{y}, \qquad \Omega_2 \, dz = \frac{z\,dz}{y} ,
\end{equation}
%
are in correspondence with the loop momenta, and turn into \cite{Gomez:2016cqb}
\begin{align*}
\left.
\left\{ \frac{dz}{y}, \, \frac{z\,dz}{y} \right\} \right|_{\l_1=a_1 \atop \l_2=a_2} \! \Rightarrow   
\left\{ \omega^1_{\sigma}\, d\sigma= \frac{ (\sigma_{1^+}-\sigma_{1^-}) \, d\sigma }{ (\sigma - \sigma_{1^+} ) (\sigma - \sigma_{1^-} ) } , \, \omega^2_{\sigma}\, d\sigma= \frac{ (\sigma_{2^+}-\sigma_{2^-}) \, d\sigma }{ (\sigma - \sigma_{2^+} ) (\sigma - \sigma_{2^-} ) } \right\}, \quad
\end{align*}
where $\sigma$ is the holomorphic coordinate on the Riemann sphere.
The four new marked points, $( \sigma_{1^+} , \sigma_{1^-} , \sigma_{2^+}, \sigma_{2^-} )$, have as momentum vectors, $ ( \ell_{1} , -\ell_{1} , \ell_{2}, -\ell_{2} ) $, in the double forward limit. Therefore, the only building blocks to construct well-defined integrands are given by the set of variables
\begin{equation}
\left\{
\frac{1}{\sigma_i - \sigma_j } , \, \omega^1_{\sigma_i} ,\omega^2_{\sigma_i}
\right\},
\end{equation}
and using a particular combination of these we can generate integrands that are quadratic in loop momentum. In \cite{Gomez:2016cqb}, a detailed analysis about this subject can be found, and it is seen that,
\begin{equation}
{\cal I}^{\rm CHY} = {\cal I} ^{\rm L} ( \omega^1) \times {\cal I} ^{\rm R} ( \omega^2)
\,\,\,\,^{ \underrightarrow{ \,\,\,  \int d\mu_n\, {\cal I}^{\rm CHY} \,\,\, }  } \,\,\,\,\,
\frac{1}{(\ell_1+\ell_2)^2 (\ell_1+\ell_2+k_1)^2 \cdots (\ell_1+\ell_2+k_1+\cdots+k_p)^2}.
\nonumber
\end{equation}

Following the same line of thought, we can generalize to two-loops, and similar ideas can be performed beyond two-loops. We consider in the two-loop case a hyperelliptic curve, $y^2=f(z)$, of degree 10, which describe Riemann surfaces of genus g=4 (it is well known that not every Riemann surface of genus g = 4 can be written as a hyperelliptic curve \cite{Harris}, however, this is not a problem if we localize on the boundary of the moduli space). On this curve, there are four global holomorphic forms that are related to the four loop momenta. Using the global residue theorem, the $A$-cycles are pinched, and the holomorphic forms turn into, 
\begin{align*}
\left.
\left\{ \frac{dz}{y}, \, \frac{z\,dz}{y},\, \frac{z^2\,dz}{y},\, \frac{z^3\,dz}{y} \right\} \right|_{{\rm Pinching} \atop {\rm A-cycles} } \! \Rightarrow   
\left\{ \omega^\a_{\sigma}\, d\sigma= \frac{ (\sigma_{\a^+}-\sigma_{\a^-}) \, d\sigma }{ (\sigma - \sigma_{\a^+} ) (\sigma - \sigma_{\a^-} ) }  \right\}, \quad \a=1,2,3,4.
\end{align*}
Thus the amplitude is localized on a Riemann sphere with eight new extra punctures, $(\sigma_{1^+},\sigma_{1^-},\ldots,\sigma_{4^+} , \sigma_{4^-} )$. The momenta associated to these marked points are in the forward limit, \ie\ $( \ell_{1} , -\ell_{1} , \ldots, \ell_{4} , -\ell_{4} )$, respectively. The results obtained in \cite{Gomez:2016cqb} suggests that CHY integrands with combinations of the products of $\{ {\rm q}_i^1, {\rm q}_i^2, {\rm q}_i^3 \}$, given in Table \ref{table},
\begin{table}[ht]
	\label{intruleTable}
	\large
	\centering
	\begin{tabular}{c|c||c|c|}
		\multicolumn{2}{c}{} & \multicolumn{2}{c}{{\small }} \tabularnewline
		\cline{2-4}
		& & 
		${\cal I}^{\rm L}$ & ${\cal I}^{\rm R}$ \tabularnewline[1ex]
		\cline{2-4}
		& \bfseries \,\,${\rm q}^1_i$\,\, &\,\, $\omega_{\sigma_i}^1 (\omega_{\sigma_i}^1-\omega_{\sigma_i}^3)$ \,\, & \,\,$\omega_{\sigma_i}^2 (\omega_{\sigma_i}^2-\omega_{\sigma_i}^4)$ \,\, \tabularnewline[1ex]
		& \bfseries \,\,${\rm q}^2_i$\,\, & \,\, $\omega_{\sigma_i}^3 (\omega_{\sigma_i}^3-\omega_{\sigma_i}^1)$ \,\, & \,\,$\omega_{\sigma_i}^4 (\omega_{\sigma_i}^4-\omega_{\sigma_i}^2)$ \,\, \tabularnewline[1ex]
		& \bfseries \,\,${\rm q}^3_i$\,\,  & \,\,$\omega_{\sigma_i}^1 \, \omega_{\sigma_i}^3$ \,\,  & \,\,$\omega_{\sigma_i}^2 \, \omega_{\sigma_i}^4$ \,\,  \tabularnewline[1ex]
		\cline{2-4}
	\end{tabular}
	\caption{Integrand building blocks.}\label{table}
\end{table}
are able to reproduce quadratic propagators in terms of the momenta, $L_1=\ell_1+\ell_2$ and $L_2=\ell_3+\ell_4$, both in the planar and nonplanar sectors. We leave following up on these ideas to future work.


\end{document}